\documentclass[12pt]{article}

\usepackage[dvipdfm]{graphicx}
\usepackage{amssymb}
\usepackage{amsmath}
\usepackage{epsfig}
\usepackage{dcolumn}
\usepackage{rotating}
\usepackage{verbatim}

\usepackage{color}
\definecolor{rosso}{cmyk}{0,1,1,0.4}
\definecolor{rossos}{cmyk}{0,1,1,0.55}
\definecolor{rossoc}{cmyk}{0,1,1,0.2}
\definecolor{blu}{cmyk}{1,1,0,0.3}
\definecolor{blus}{cmyk}{1,1,0,0.6}
\definecolor{bluc}{cmyk}{1,1,0,0.1}
\definecolor{verde}{cmyk}{0.92,0,0.59,0.25}
\definecolor{verdec}{cmyk}{0.92,0,0.59,0.15}
\definecolor{verdes}{cmyk}{0.92,0,0.59,0.7}

\newcommand{\ba}{\begin{eqnarray}}
\newcommand{\ea}{\end{eqnarray}}
\newcommand{\be}{\begin{equation}}
\newcommand{\ee}{\end{equation}}
\newcommand{\al}{\alpha}

\newcommand{\da}{\delta}
\newcommand{\la}{\lambda}

\newcommand{\sa}{\sigma}
\newcommand{\en}{\epsilon}
\newcommand{\oa}{\omega}
\newcommand{\Ga}{\Gamma}

\newcommand{\La}{\Lambda}
\newcommand{\cO}{{\cal O}}

\newcommand{\cK}{{\cal K}}

\newcommand{\w}{\widetilde}

\newcommand{\p}{\partial}

\newcommand{\ra}{\rightarrow}
\newcommand{\Ra}{\Rightarrow}

\newcommand{\LF}{\left(}
\newcommand{\RF}{\right)}
\newcommand{\LT}{\left[}
\newcommand{\RT}{\right]}

\newcommand{\2}{\frac{1}{2}}

\newcommand{\stwo}{\sqrt{2}}

\newcommand{\mx}{\mbox}
\newcommand{\mt}{\mathtt}
\newcommand{\mand}{\mx{ and }}

\newcommand{\ie}{{\it i.e.\ }}

\newcommand{\Ma}{M_{\ast}}
\newcommand{\sech}{\mt{sech}}
\begin{document}
\tolerance=100000
\thispagestyle{empty}
\vspace{1cm}

\begin{center}
{\LARGE \bf
Towards a Resolution of the Cosmological Singularity in Non-local Higher Derivative Theories of Gravity
 \\ [0.15cm]
% \&
% \\ [0.18cm]
% when it is long
}
\vskip 2cm
{\large Tirthabir Biswas}~$^{a,~b}$,~
{\large Tomi Koivisto}~$^{c}$,~and
{\large Anupam Mazumdar}~$^{d, ~e}$
\vskip 7mm
{\it $^a$~Department of Physics,
St. Cloud State University, St. Cloud, MN 56301}
{\it $^b$~Physics Department,
Loyola University, Campus Box 92, New Orleans, LA 70118}
\vskip 3mm
{\it $^c$~Institute for Theoretical Physics and Spinoza Institute, Postbus 80.195, 3508 TD Utrecht, The Netherlands}
\vskip 3mm
{\it $^d$~Physics Department, Lancaster University, Lancaster, LA1 4YB, United Kingdom}\\
{\it $^e$~Niels Bohr Institute, Blegdamsvej-17, Copenhagen-2100, Denmark}
\end{center}
\date{\today}

\begin{abstract}
One of the greatest problems of standard cosmology is the Big Bang singularity. Previously it has been shown
that non-local ghostfree higher-derivative modifications of Einstein gravity in the ultra-violet regime can admit non-singular bouncing solutions. In this paper we study in more details the dynamical properties of the equations of motion for these  theories of gravity in presence of positive and negative cosmological constants and radiation. We find stable inflationary attractor solutions in the presence of a positive cosmological constant which renders inflation {\it geodesically complete}, while in the presence of a negative cosmological constant a cyclic universe emerges. We also
provide an algorithm for tracking the super-Hubble perturbations during the bounce and show that the bouncing solutions are  free from any perturbative instability.

\end{abstract}

\newpage

\setcounter{page}{1}

\tableofcontents
%%%%%%%%%%%%%%%%%%%%%%%%%%%%%%%%%%%%%%%%%%%%%%%%%%%%%

\section{Introduction}

Primordial {\it inflation} is one of the most well motivated paradigms of the early universe, for a recent review, see~\cite{MR}.
A large e-foldings of inflation can explain the large scale structures of the universe and also the origin of seed perturbations for the cosmic microwave background radiation. In spite of the great successes of inflation, it does not address one
of the most important aspects of the Big Bang cosmology. For any
equation of state obeying the strong energy condition $p>-\rho/3$,
regardless of the geometry (flat, open, closed) of the universe, the
scale factor of the universe in a Friedmann Robertson Walker (FRW)
metric vanishes at $t=0$, and the matter density diverges. In fact all
the curvature invariants, such as $R,~\Box R,...$, become singular.
This is the reason why it is called the {\it Big Bang singularity
problem}.

Although inflation requires $p<-\rho/3$, it does not alleviate the Big
Bang singularity problem, rather it pushes the singularity backwards
in time. Many authors have pondered on this debatable issue, whether
inflation is past eternal or not~\cite{Borde,Guth,Linde}, and the
conclusion is that it is not, at least in the context of Einstein
gravity as long as the average expansion rate in the past is greater
than zero, i.e. $H_{av}>0$~\cite{Guth}. The fluctuations grow as the
universe approaches the singularity, and the standard {\it singularity
theorems} due to Hawking and Penrose hold, which inevitably leads to a
collapse in FRW geometry as long as the energy
density is positive~\cite{Hawking} (see also~\cite{Guth}).

There have been many attempts to resolve this cosmological singularity problem
classically~\cite{Hawking}, and in the context of string theory~\cite{Many}, for
a review see~\cite{Craps}, but none has successfully resolved the space-like
singularity in the context of a flat universe. All the stringy toy models which have
been constructed so far seem to suffer from classical and/or quantum instabilities and inconsistencies,
%in the presence of time dependent background
 such as the presence of negative tension branes, closed time like curves, negative-norm ghost states, etc., see~\cite{Cornalba,Horowitz,Seiberg}
for details.

One alternative view would be to study fully non-perturbative and ghostfree higher derivative actions of gravity inspired from string  theory~\cite{BMS}: An infinite set of higher derivative corrections (series expansion in $\al'$, the string tension) mostly in the form of an exponential are often known to appear in stringy models, such as tachyonic actions in string field
theory~\cite{sft1}-\cite{pressure} (for a review
see~\cite{sft_review}),  Bulk fields localized on a codimension-2
branes \cite{claudia} and various toy models of string theory such
as $p$-adic strings~\cite{padic_st} (see also~\cite{zwiebach}), zeta
strings~\cite{zeta}, and strings quantized on a random lattice
\cite{random,marc} (see also \cite{ghoshal}).
These corrections appear already at classical level
(i.e., at the tree level), as well as  loop level or even non-perturbatively (e.g., as expected in the 1/N expansion in some Yang-Mills theory~\cite{Hooft}). Incidentally, recently it was shown~\cite{bck} how one can consistently perform quantum loop calculations in these non-local models  and  recover, in the context of finite temperature field theory, some of the novel properties of string theory such as thermal duality and Hagedorn transition. Thus it is interesting to explore whether similar theories of gravity, albeit only at the level of toy models, can provide us insights into the resolution of cosmological singularities.

It is worth while to point out that like most higher derivative actions, these theories have improved Ultra-violet (UV) behavior and  can render gravity asymptotically free, but importantly without introducing ghost states~\cite{BMS}, for some earlier attempts, see~\cite{Stelle,Van,Zwiebach}, which were not very successful.  Note that asymptotic freedom  is required (in addition to renormalizability) in ordinary field theory for phenomenological reasons (scaling in processes such as deep
inelastic scattering), as well as to avoid certain problems in the nonperturbative definition of the theory. This condition usually takes the form of ``asymptotic safety" \cite{Weinberg}, a more general requirement
for having a well-defined UV limit. The theories we consider are therefore natural candidates for such GR extensions. Building on the work done in~\cite{BMS}, in this paper we continue to study cosmological  background solutions and their perturbations arising in the  higher derivative nonlocal gravity models, and in particular whether  they can resolve the Big Crunch/Bang singularity problem of GR robustly. Previously, non-local~\footnote{Some of these models are very similar to the models we consider in our paper and contain and infinite series of higher derivative terms, but in the literature highly non-local operators, such as $f(1/\Box)$, have also been considered.}   higher derivative corrections to gravity has been studied in the context of black hole singularities~\cite{warren}, understanding ultra-violet behavior~\cite{uv-nonlocal},  cosmological applications (for instance, involving  dark energy~\cite{Woodard},  inflation~\cite{inflation}, and  generating
density perturbations~\cite{pert}), and modifications to the Newtonian potentials in the metric~\cite{nonlocal-gravity}. For discussions on these topics in the more general context of higher derivative theories of gravity, see~\cite{gravity}~\footnote{In parallel, string-inspired non-local scalar field theories have also been studied quite extensively in recent times, including  several interesting applications~\cite{scalar}.}.

In~\cite{BMS}  non-local asymptotically safe theories of gravity were already constructed and it was shown that they can admit non-singular bouncing solutions. However, it was not clear how generic these solutions were, whether these solutions were stable under small perturbations, and whether these theories contain other singular solutions or not. The primary focus of this paper is to shed some light on these issues, and also see whether they can provide phenomenologically viable cosmological models. For instance, can one have a geodesically complete inflationary paradigm once the non-local modifications to gravity are included? We know that for the success of the inflationary paradigm,  the super-Hubble fluctuations must ``freeze'', is that true in the non-local models or do the fluctuations decay/ grow  during inflation? In cyclic and bouncing alternatives~\cite{narlikar}-\cite{prebigbang}, the issue of transmitting perturbations is of paramount importance. Can we understand the evolution of the fluctuations in the concrete nonsingular bouncing set-up that our model provides?

In this paper specifically we will seek a two parameter family of solutions, labeled by $\lambda$ and $\mu$, which solves the higher derivative theory of gravity. Interestingly, this class of solutions demands the presence of a cosmological constant, which is no longer free but is determined in terms of the two parameters, and  radiation. One particular solution for the scale factor turns out to be a {\it hyperbolic cosine } bounce with a positive cosmological constant, where $\mu=-6\lambda^2$~\cite{BMS}.
In principle $\lambda$ can be both positive and negative in order to provide a non-singular bouncing cosmology, but
for $\lambda < 0$, it is not possible to recover the GR limit at late times. The most interesting case is when $\lambda >0$ where
not only we recover the right infrared behavior of gravity, but also two interesting classes of solutions with positive and negative
cosmological constants, i.e. $\pm \Lambda$.

The case of $\Lambda >0$ is very interesting from the point of view of inflation, as it paves the way for a {\it geodesically}
complete paradigm in past and future. There exists a deSitter solution as an attractor for large enough times before and after the bounce. This is a good news for both low scale~\cite{AEGJM} and an intermediate scale models of inflation~\cite{EMS}
which are driven by the vacuum energy of the inflaton field which can excite the observed relativistic and non-relativistic degrees of freedom after the end of inflation. On the other hand the $\mu>0,\Lambda <0$ case is equally interesting from the point of view of providing an early phase of  cyclic cosmology. In this case there does not exist any deSitter attractor but there are also no
singular solutions. Although, a mere cyclic phase is a rather boring scenario as it fails to reproduce the universe
we live in,  recently it has been proposed that once interactions between radiation and non-relativistic matter are taken into account then we can obtain a cyclic inflationary paradigm where the  cycles become asymmetric and the universe expands bit by bit  to give rise to an exponential inflation over many many cycles~\cite{BM}, see also~\cite{AB}. Both $\Lambda>0, \Lambda<0$ cases will be studied analytically and numerically in our paper.

In general, taking into account a preceding contracting phase removes the inflationary infrared divergences of quantum fluctuations \cite{Koivisto:2010pj}.
Here we study the evolution of the super-Hubble perturbations during the bounce and ensure that they do not feature any classical and/or quantum instabilities. Interestingly, we find that on large scales all the modes are decaying except one which remains frozen. This frozen mode could be useful if we wish to evolve the perturbations either prior or after the bounce in order to match the observed CMB perturbations. However, we will not delve into any details about these issues in this paper.

We will begin our discussion with a brief recap of nonperturbative  higher derivative action and its equation of motion in section 2. We will discuss the background solutions in section 3, where we also provide an analytical example of a specific class of non-local models. In section 4, we will study various cases of $\lambda >0, \lambda <0$  and their dynamical attractor solutions for $\Lambda >0,\Lambda <0$. In section 5, we track the perturbations during the bounce. Some series expansions are confined to the appendices: \ref{app1} presents a general series solution for the two-parameter ansatz, and \ref{app2}, \ref{app4} \ref{app3} consider some special cases. We conclude in section 6.

%%%%%%%%%%%%%%%%%%%%%%%%%%%%%%%%%%%%%%%%%%%%%%%%%%%%%%%%%%%%%%

\section{Ghost-free non-local theories of gravity}

\subsection{Ghost-free higher derivative actions}
\label{ghostfree}

In \cite{BMS}, a string-inspired ghost-free and asymptotically free higher derivative
gravity action was introduced. In this paper we investigate their cosmology in more depth and generality, and suggest to use this action to study how perturbations propagate across the bounce.  Before doing this, we shall review the construction of \cite{BMS}.

The action discussed in \cite{BMS} takes the form
\be \label{action}
S \, = \, \frac{M_p^2}{2}\int d^4x\ \sqrt{-g}F(R) \, ,
\ee
where $M_p$ is the reduced Planck mass and
\be
F(R) \, = \, R \, + \,
\sum_{n=0}^{\infty}{c_{n}\over M_{\ast}^{2(n + 1)}}R\Box^{n}R \, .
\label{non-pert}
\ee
Here, $M_{\ast}$ is the mass scale at which the higher derivative
terms in the action become important. In order to ensure that
the action is ghost-free, it is important that the series does
not truncate at a finite order, i.e. the coefficients $c_n$ must
remain non-vanishing to arbitrarily high orders. For the purpose of illustration, let us provide a specific example:
If we choose
\be \label{coeffs}
c_n \, = \, - {1 \over 6} {{(-1)^{n + 1}} \over {(n + 1)!}} \, ,
\ee
 the series can be summed up and $F(R)$ becomes
\be
F(R) \, = \, R - {R\over 6}
\bigl( {{ e^{- {\Box} / {M_{\ast}^2}} - 1} \over {\Box}} \bigr) R \, .
\ee

To understand the absence of ghosts, it is convenient to re-write
the action in terms of a scalar field $\psi$ and a  Lagrange multiplier
field $\phi$ in the form
\ba
S \, &=& \, {M_p^2\over 2}\int d^4x \sqrt{-g} \LF \Phi R +
\psi \sum_{n = 1}^{\infty} c_n {{{\Box}^n}\psi \over {M_{\ast}^{2(n + 1)}}} - \bigl[ \psi (\Phi - 1) - {{c_0} \over {M_{\ast}^2}}\psi^2 \bigr] \RF \, .
\ea
After performing conformal transformations, $g_{mn}\ra  \Phi g_{mn}$, and linearizing around $\Phi=1$ we find
\ba
S \, &=& \, {M_p^2\over 2}\int d^4x \sqrt{-g} \LF  R + {3\over 2}\phi\Box\phi
+\psi \sum_{n = 0}^{\infty} c_n {\Box^n\psi \over {M_{\ast}^{2(n + 1)}}} - \psi \phi  \RF \, .
\ea
where $\Phi=e^{\phi}\approx 1+\phi$. We note in passing that while $\phi$ is dimensionless, $\psi$ has mass dimension 2.
The variational equations with respect to $\phi$ and $\psi$ yield
\be
\psi \, = \, 3 \Box \phi \,
\ee
and
\be
\phi \, = \, 2 \sum_{n = 0}^{\infty}c_n {\Box^n\psi \over M_{\ast}^{2(n + 1)}}  \, ,
\ee
respectively. We can now combine these equations to obtain
\be \label{phieq}
\LT 1 - 6 \sum_{n = 0}^{\infty} c_n
{{\Box^{n + 1}} \over {M_{\ast}^{2(n + 1)}}} \RT \phi \,
\equiv \, \Gamma(\Box) \phi \, = \, 0 \, .
\ee
Ghosts would appear as poles of the propagator of $\phi$. If we can
sum up the infinite series on the right hand side of (\ref{phieq})
into the form
\be \label{gammas}
\Gamma(\Box) \, = \, e^{\gamma(\Box)} \, ,\mx{ or } \Gamma(\Box) \, = \, \LF1-{\Box\over m^2}\RF e^{\gamma(\Box)}
\ee
where $\gamma(w)$ is an analytic function in the entire complex
plane, then it follows~\cite{BMS} that there are no ghosts. One
simple example of a ghost-free action results if we choose
\be
\gamma(w) \, = \, w \, ,
\ee
which is precisely what one obtains when the coefficients $c_n$ are chosen as in (\ref{coeffs}). In case one is concerned that integrating out the field $\psi$ above changes the conclusions, we present alternative derivations of the ghost-free condition in the appendix \ref{ghost_app}.

One way to see that the action given by (\ref{action}) with the
coefficients chosen as in (\ref{coeffs}) is asymptotically free
is to solve \cite{BMS} the linearized field equations for a point
source and to observe that the contribution of ultraviolet modes is
exponentially cut off. This is a rather typical scenario in these models, the UV behavior improves due to the exponential ``regulation''.

The asymptotic freedom of the theory leads to nonsingular cosmologies.
An exact nonsingular solution was constructed in \cite{BMS} for a limited class of non-local actions of the form (\ref{non-pert}). One of the focus of this paper is to generalize the class of bouncing solutions, so that we can look at a much larger set of non-local actions. Also we want to study the stability and phenomenological viability of these models.
%The main aim of this section is to generalize the procedure so that in particular we can study the case when the cosmological constant is zero and also generalize the analysis to study perturbations.

%%%%%%%%%%%%%%%%%%%%%%%%%%%%%%%%

\subsection{Einstein-Schmidt (ES) equations}
For  this purpose we first want to simplify the modified Einstein's equations for our action (\ref{non-pert}). We will henceforth refer to them as ES equations, as much of the work on such generalizations was done by Schmidt~\cite{Schmidt}. The ES equations look like
\be \label{e-s}
\w{G}_{\mu\nu}=\frac{1}{M_p^2}T_{\mu\nu}\,,
\ee
where $T_{\mu\nu}$ is the matter stress tensor and the generalized Einstein tensor $\w{G}_{\mu\nu}$, called here the ES tensor, has the form
\ba
\w{G}_{\mu\nu} & = & F_0R_{\mu\nu}-\2Fg_{\mu\nu}-F_{0;\mu\nu}+g_{\mu\nu}\Box F_0 \nonumber \\ & + & \2\sum_1^{\infty}\LT
g_{\mu\nu}(F_N(\Box^{N-1}R)^{;\sa})_{;\sa}-F_{N;(\mu}(\Box^{N-1}R)_{;\nu)}\RT\,,
\ea
%{\bf I changed the sign of the 2nd term. -tomi}
where we have defined
\be
F_N=\sum_{M=N}^{\infty}\Box^{M-N}\LF{\p F\over \p \Box^MR}\RF\,,
\label{Fn}
\ee
and the symmetrization convention is such that $A_{(\mu\nu)} \equiv A_{\mu\nu} + A_{\nu\mu}$.
Let us obtain the ES tensor, $\w{G}_{\mu\nu}$, for the special case when
\be
F=R\Box^nR\,.
\ee
One finds
\ba
\w{G}^n_{\mu\nu}=2R_{\mu\nu}\Box^nR-2(\Box^n R)_{;\mu\nu}+2g_{\mu\nu}\LF\Box^{n+1}R-\4R\Box^nR\RF \nonumber \\
+\2\sum_{m=1}^{n}\LT g_{\mu\nu}(\Box^{n-m}R(\Box^{m-1}R)^{;\sa})_{;\sa}-(\Box^{n-m}R)_{;(\mu}(\Box^{m-1}R)_{;\nu)}\RT\,.
\ea
Since the ES tensor is linear in $F$, the complete ES tensor for the more general F in Eq.(\ref{non-pert}) is given by
\be
\w{G}_{\mu\nu}=G_{\mu\nu}+\sum_{n=0}^{\infty}{c_n\over \Ma^{2(n+1)}}\w{G}^n_{\mu\nu}
\ee

In particular it is rather useful to look at the trace of the ES tensor. Defining as before
\be \label{e-s_trace}
\w{G}\equiv g^{\mu\nu}\w{G}_{\mu\nu}=G+\sum_{n=0}^{\infty}{c_n\over \Ma^{2(n+1)}}\w{G}^n\,,
\ee
where
\be
\w{G}^n=6\Box^{n+1}R+\sum_{m=1}^{n}\LT (\Box^{n-m}R)_{;\mu}(\Box^{m-1}R)^{;\mu}+2(\Box^{n-m}R)(\Box^{m}R)\RT
\ee
gives the contribution to the trace of the ES tensor from each derivative order of $(n+4)$.

%%%%%%%%%%%%%%%%%
\section{Finding bouncing backgrounds}
\subsection{The ansatz}
Quite remarkably infinite differential equations similar to what we are interested in has been studied by mathematicians, see
\cite{Barnaby}, and several rigorous results exist, especially for linear systems. The various techniques and results developed has now been successfully applied to analyze cosmology involving infinite differential equations involving scalar fields as one encounters in p-adic~\cite{p-adic} and string field theory~\cite{sft1}-\cite{pressure}. In particular what has been realized is that mostly the non-local infinite differential equations can be approximately mapped to a collection of local field theories. To make things concrete, consider a non-local action of the form
\be
S=M^2\int d^4x \phi\Ga(\Box)\phi
\ee
leading to the field equation
\be
\Ga(\Box)\phi=0\,.
\label{free-scalar}
\ee
Let us assume that $\Ga$ is an analytic function with $N$ distinct zeroes of order one, at points $m^2_i$ with $i=1\dots N$. Then one can rigorously show that the set of classical trajectories for the full non-local action is simply given by the ``union" of solutions to the following ``local'' field theory equations
\be
(\Box+m_i^2)\phi=0\ ,\ i=1\dots N
\ee
One can thus use this correspondence to obtain solutions to the full non-local equations by solving only the usual field theory equations. When one adds interactions to the theory, then such a correspondence is no longer exact, but again one can show that one can obtain approximate solutions to the original non-local (and now non-linear) equations from local field theory solutions. This procedure was successfully used to obtain, not only background solutions but also in computing the spectrum of perturbations around the background~\cite{p-adic}. Our aim in this section is to develop a similar algorithm to obtain solutions (both background and perturbations) for our non-local theory of gravity.

We are going to focus on the situations where there is only a cosmological constant and radiation as the matter source. We will see that this situation  is similar to the  scalar field with no interaction term (\ref{free-scalar}). Thus one expects to be able to find exact solutions of the non-local theory by solving only ``local'' field equations. There is however one crucial difference:   gravity being a gauge theory  leads to additional constraints which are different from the  ``local'' analogues. This explains why we can get qualitatively very different background solutions in the non-local theories as compared to the local counterparts. This will become clearer as we proceed. To begin with let us provide the ``local ansatz'' for our non-local theories.

%We have so far described our nonlocal gravity sector. For the matter content, we will consider  radiation along with a cosmological constant.
In order to implement the mapping between non-local and local theory, it is imperative that we start with the trace equation for gravity $M_p^2\w{G}=T$, where $\w{G}$ is given by (\ref{e-s_trace}) and $T$ is the trace of the matter stress tensor:
\ba
\w{G} & = & -R+\sum_0^{\infty}\frac{c_n}{\Ma^{2(n+1)}}\LF 6\Box^{n+1}R+\sum_{m=1}^{n}\LT
(\Box^{n-m}R)_{;\mu}(\Box^{m-1}R)^{;\mu}+2(\Box^{n-m}R)(\Box^{m}R)\RT\RF \nonumber \\ & = &
-4{\La\over
M_p^2} \,.
\label{trace}
\ea
%{\bf Just added the scales $\Ma$ -tomi}.
Since radiation is conformal, it doesn't contribute to the trace equation.
In close analogy with the scalar case we now decide to look at the subset of solutions which satisfies the ansatz
\be \label{ansatz}
\Box R=\la R + \mu\,,
\ee
where $\la$ has the dimensions of $M^2$ and $\mu$ has the dimensions of energy density, $M^4$.
Unlike the scalar case we do not however claim that the subset of solutions of all the ansatz of the above form gives us the complete set of solutions to the full non-local equations. For the purpose and scope of this paper we will restrict ourselves to solutions of the non-local equations which also satisfy the above ansatz.

Now, from (\ref{ansatz}) it follows that
\be
\Box^nR=\la^{n}R+\la^{n-1}\mu \,.
\label{recursion}
\ee
One can simplify the terms in the right hand side of (\ref{trace}), and after some straight forward but tedious algebra one finds
\be
\w{G}=A_1 R+A_2\LF R^2 - {\dot{R}^2\over2\la}\RF + A_3\,,
\ee
%{\bf I added the minus sign before $\dot{R}^2$ -tomi}
where
\ba
A_1&=&-1+6\sum_{n=0}^{\infty}c_n\LF{\la\over \Ma^2}\RF^{n+1} + 2\mu\sum_{n=1}^{\infty}(2n-1)
\frac{c_n}{\Ma^4}\LF{\la\over \Ma^2}\RF^{n-1}  \nonumber \\
& = & -\Ga(\la) - 2\frac{\mu}{\lambda^2}\LF
\frac{1}{6}\left(1-\Ga(\la)\right)-\frac{c_0\lambda}{\Ma^2}\RF + \frac{2\mu}{\la}A_2 \,, \\
A_2 &=& 2\sum_{n=1}^{\infty} \frac{nc_n}{\Ma^2}\LF{\la\over \Ma^2}\RF^{n} \,, \\
A_3 &=& 6\mu\sum_{n=0}^\infty {c_n\over \Ma^2} \LF\frac{\la}{\Ma^2}\RF^{n} +
2\mu^2\sum_{n=2}^{\infty}(n-1){c_n\over
\Ma^6}\LF\frac{\la}{\Ma^2}\RF^{n-2} \nonumber
\\ & = & \frac{\mu}{\la}\LF 1-\frac{\mu}{3\lambda^2}\RF\LF 1-\Ga(\la)\RF +
\frac{\mu^2}{\la^2}\LF 2\frac{c_0}{\Ma^2}+A_2 \RF \,.
\ea
%{\bf added the $\sim \mu$ terms and changed a power of $\Ma$ in $A_2$. -tomi}
We can now equate the coefficients of $R$ and the square of $R$ to zero and require the
remaining constant to be given by the $\Lambda$, to obtain solutions to the trace equation.
This implies three conditions, which can be put into the following simple form
\ba
\Ga(\la) & = & \frac{-1+6c_0\la/M_*^2}{3\la^2-\mu}\mu \,, \label{tr-cons1}\\
\Ga'(\la) & = & \frac{\Ga-1}{ \la} \,, \label{tr-cons2} \\
\Lambda  & = & -\frac{\mu M_p^2}{4\la} \,.
\label{tr-cons3}
\ea
Thus given a specific higher derivative gravity action ($c_n$'s), we can find an ansatz of the form (\ref{ansatz}) which solves the trace equation
corresponding to the non-local action (\ref{action}). (\ref{tr-cons2}) determines the value of $\la$, and (\ref{tr-cons1}) then specifies  the value of $\mu$ uniquely. However, one has to still satisfy the additional constraint (\ref{tr-cons3}).   Due to our specific ansatz, the cosmological constant can no longer be a free parameter, but instead is given by
the equation (\ref{tr-cons3}). Another way of thinking about this is that given a specific cosmological constant, we can only find solutions of the
type (\ref{ansatz}) for a subclass of non-local actions under consideration. In section \ref{cosh_sec} we will provide an illustrative example of a class of non-local models admitting bouncing solutions.

%%%%%%%%%%%%%%%%%%%%%%%%%%%%%%%%%%%%%%%%%
\subsection{Nonsingular bounces}
Since we have ``reduced'' the non-local infinite differential equations to a finite fourth order equation (\ref{ansatz}), this is not a very difficult problem to analyze and can be done rigorously  for instance, using power series methods. For simplicity we will focus on symmetric bouncing solutions, so that the Hubble constant admits an expansion like the following
\be
H=\Ma\sum_{n=0}^{\infty}h_{2n+1}(\Ma t)^{2n+1}\,.
\label{H-expansion}
\ee
By substituting (\ref{H-expansion}) in (\ref{ansatz}) one can now systematically determine  the $h_n$'s by looking order by order in $(\Ma t)^2$. Actually, since (\ref{ansatz}) is a third order differential equation in $H$, we expect three independent ``integration constants''. By choosing a symmetric bouncing solution (\ref{H-expansion}) we have implicitly chosen $h_0=h_2=0$. Thus we are still left with one unconstrained parameter in $h_1$.
%As an illustration of this procedure by looking at the zeroeth order (constant) term in (\ref{ansatz}) one finds {\bf I got other $h_3$ %-Tomi}
%\be
%h_3=-{1\over 3}\LT{\la^2(\mu^2+6\Ma^2)\over 12 \Ma^4}+2\RT
%\ee
%As mentioned before, $h_3$ and indeed all $h_n$'s are uniquely determined in terms of $\mu^2,\la^2$ and $\Ma^2$. We already know that the first two parameters are determined from the non-local action, but what determines $\Ma$? One may also wonder about the usual $R^2$ theory, whose trace equation has precisely the form of the ansatz (\ref{ansatz}). Does it mean that ordinary $R^2$ gravity admits non-singular bouncing solutions?

In appendix \ref{app1} we have derived the series expansion and in appendices \ref{app2} and \ref{app3} considered some special cases. The first few $h_n$'s are:
\ba
h_3 & = & -\frac{1}{36\Ma^4}\LF\mu + 6h_1\Ma^2(\la+4h_1\Ma^2) \RF\,, \label{hh_3} \\
h_5 & = & \frac{1}{720\Ma^6}\LF \la\mu + 2h_1(3\la^2+11\mu)\Ma^2+132\la h_1^2\Ma^4 + 384h_1^3\Ma^6 \RF\,, \\
h_7 & = & -\frac{1}{90720 \Ma^8}\Big[ \mu (3 \la^2 + 26 \mu) + 6 h_1 \la (3 \la^2 + 73 \mu) \Ma^2 \nonumber \\
& + &  12 h_1^2 (141 \la^2 + 242 \mu) \Ma^4 + 17424\la h_1^3 \Ma^6 + 39168 h_1^4 \Ma^8\Big]\,. \label{hh_7}
\ea
From the structure of the series it is apparent that each coefficient $h_{2n+1}$ is uniquely determined  by the previous ones. The highest order coefficient appears always linearly, guaranteeing a real solution. Thus given a specific non-local model (and the appropriately  fine-tuned cosmological constant) we have a one-parameter solution of symmetric bounces, the parameter being $h_1$, or equivalently, the value of $\dot{H}$ at the bounce point. An important point to note is that although $\dot{H}\propto h_1$ and hence can be as small as one wants, the time scale of the bounce is still naturally governed by the $\Ma$ scale. As one can see from (\ref{hh_3}-\ref{hh_7}) all the higher order $h_n$'s will  naturally be $\cO(1)$, even if $h_1\ra 0$.
%%%%%%%%%%%%%%%%%%%%%
\subsection{Bianchi Identities \& the energy density at the bounce}
So far we have only solved the trace equation, and one may wonder whether this is sufficient.
The answer is yes,  and has to do with the fact that gravity is a constrained system and satisfies the Bianchi identity. For time dependent solutions
\be
\w{G}=-4{\La\over M_p^2}\Ra \w{G}_{00}=\frac{\rho_0}{M_p^2} a^{-4}\,.
\ee
In other words, the modified Hubble equation can (at most) admit radiation which being conformal is traceless and therefore doesn't show up in the trace equation. Thus the solutions that we obtain via our ``local'' ansatz provides a solution to the complete set of ES equations (\ref{e-s}), but in so doing may inherently assume the presence of some amount of radiation as well. Although, there is nothing wrong with having a bit of radiation, we need to determine $\rho_0$ because $\rho_0$ may turn out to be negative. This would mean having a ghost like radiation degree of freedom, precisely the kind of states we set out  to avoid.
%In fact, for the pure $R^2$ gravity case, this is what happens. Thus there is no conflict with what people already know, $R^2$ gravity cannot solve the singularity problem. Second, one of the purpose of this paper is to look at perturbations. When we are looking at perturbations of a specific background solution with a given background radiation, we must make sure that the perturbed equations correspond to the same background radiation. In other words some of the perturbation modes which solves (\ref{ansatz}) may be spurious and needs to be projected out. Finally, one can use $\rho_0$ to determine the scale of the bounce, the $\Ma$ parameter. So, for instance if there is no radiation, the equation $\rho_0=0$ should  determine $\Ma$, and in fact also tell us whether our non-local theory  admits bouncing solutions.

In order to compute $\rho_0$ and $p$ one has to look at the $\w{G}_{00}$ and $\w{G}_{ii}$ equations:
\ba
\w{G}_{00} & = & F_0R_{00}+\2F-\ddot{F_0}-\Box F_0-\2\sum_1^{\infty}\LT F_n\Box^nR+\dot{F_n}\dot{(\Box^{n-1}R)}\RT\,,\\\label{hubble}
\w{G}_{ii} & = & F_0R_{ii}-\frac{a^2}{2} F-F_{0;ii}+a^2\Box F_0 + \frac{a^2}{2}\sum_1^{\infty}\LT F_n\Box^nR-\dot{F_n}\dot{(\Box^{n-1}R)}\RT\,.
\ea %The last coefficient of the Hubble eq. had an extra factor 3, it is removed %
where $F_n$'s are defined via (\ref{Fn}). For our specific action this simplifies to
\ba
F_n &=& \sum_{m=n}^{\infty}c_m\Box^{m-n}R \ \forall n\neq 0 \nonumber\\
F_0 &=&2\sum_{m=0}^{\infty}c_m\Box^{m}R  \mx{ for } n=0\,.
\ea
For the energy density and pressure we then get, using the constraints (\ref{tr-cons1})-(\ref{tr-cons3}) from the trace equation:
\ba
\frac{\rho}{M_p^2} & = & 9\frac{\la-2c_0\mu/\Ma^2}{3\la^2-\mu}\left[ H\left(H\la+2\ddot{H}\right)+6H^2\dot{H}-\dot{H}^2\right] +
\frac{1-6c_0\la/\Ma^2}{12\la^3-4\la\mu}\mu^2\,, \label{rho1} \\
p & = & \frac{1}{3}\rho+\frac{\mu M_p^2}{3\la} \,.
%p & = & \frac{3\left(H(H\la+2\ddot{H})+6H^2\dot{H}-\dot{H}^2\right)\left(\la-2c_0\mu/\Ma^2\right)}{3\la^2-\mu} \\ \nonumber
%& + & \frac{7\la^2-2(1+c_0\la/\Ma^2)\mu}{12\la^3-4\la\mu}\mu \,.
\ea
consistent with having  the cosmological constant and radiation.
It is easy to check that the above two equations imply trace equation. Another consistency check is to compute the continuity equation,
\be
\dot{\rho}+3H(\rho+p) = 0,
\ee
which now indeed holds identically.
We note that the solutions do not depend on the details of the higher order contributions to theory once the constraints
(\ref{tr-cons1})-(\ref{tr-cons3}) are satisfied. Namely, the coefficients $c_n$'s do not enter explicitly into the expressions, but
their sums have to satisfy the three conditions.
%\footnote{If we imposed $\Ga=1$ instead of the constraint (\ref{tr-cons1}), we would recover the general relativistic expressions with a constant contribution from $c_0$.}.
Only the value of $c_0$ is important, though only when the cosmological
constant is nonzero.

The expression for $\rho_0$, the radiation energy density at the bounce point can be obtained as follows: Notice because of our ansatz, at the bounce point $H=\ddot{H}=0$ and $\dot{H}=h_1\Ma^2$. Thus we have:
\be
\rho_0 = \frac{3M_p^2 (\la - 2c_0\mu/\Ma^2) (\mu - 12 h_1^2 \Ma^4)}{12\la^2 - 4 \mu}\,.
\label{rhonot}
\ee

Again we find that as long as $\mu\neq 0$, then even if $h_1\ra 0$, ``naturally'' we expect $\rho_0\sim\cO(M_p^2\Ma^2)$. To have a consistent bouncing solution we must make sure that $\rho_0>0$.

%%%%%%%%%%%%%%%%%%%%%%%
\subsection{An analytic example: hyperbolic cosine bounce \& a specific class of non-local models} \label{cosh_sec}
We have seen so far how we can find bouncing solutions given a non-local action. For the purpose of illustration, let us now look at a specific analytic example of a bouncing solution which obeys the ansatz (\ref{ansatz}) and was already discussed in \cite{BMS}:
\be \label{cosh}
a(t) = a_0\cosh{\sqrt{\frac{\la}{2}}t}\,,
\ee
where $\la>0$. This background satisfies our ansatz with the specific parameter combination
\be
\Box R = \la R-6\la^2\,\Ra \mu=-6\la^2\,.
\ee

The general results (\ref{tr-cons3}) and (\ref{rhonot})  imply that we need a cosmological constant, and in general also radiation to support this solution
\ba
\Lambda & = & \frac{3}{2}\la M^2_{pl}\,, \\
\rho_0 & = & -\frac{M^2_{pl}}{2\la}(\la^2+2h_1^2\Ma^4)(1+12\la c_0/\Ma^2)\,.
\ea
The last equality implies we need a negative quadratic coefficient $c_0 < -\Ma^2/(12\la)$. In our expansion for the Hubble parameter, if we choose $\oa=\sqrt{\la}$ and $h_1=1/2$, the
general solution (\ref{hh_3}-\ref{hh_7}) give us
\be
H(t) = \frac{1}{2}\la t-\frac{1}{12}\la^2t^3+\frac{1}{60}\la^3t^5-\frac{17}{5040}\la^4 t^7 + \dots\,,
\ee
which is nothing but the expansion corresponding to  (\ref{cosh}). We have thus shown that consistent hyperbolic cosine bounces with parameter $\la>0$ exist in models satisfying the following
requirements:
\ba
\Ga(\la) & = & \frac{2}{3}(1-6c_0\la/M_*^2) \,, \label{tr-cons1b}\\
\Ga'(\la) & = & \frac{\Ga-1}{ \la} \,, \label{tr-cons2b} \\
c_0 & < & -\frac{\Ma^2}{12\la}
%\Lambda  & = & -\frac{\mu M_p^2}{4\la} \,.
\label{tr-cons3b}
\ea

Clearly, $\la$ is over-determined, but there are infinite number of different sets of $c_n$'s which can satisfy the above constraints and therefore admits the background evolution (\ref{cosh}). For the purpose of illustration, let us provide a one-parameter family labeled by $\Ga_0$  of non-local actions which satisfies (\ref{tr-cons1b}-\ref{tr-cons3b}). Consider the following $\Ga(\Box)$:
\ba
\Ga(\Box)&=&\exp\LT\al_1\LF{\Box\over \Ma^2}\RF+\al_2\LF{\Box\over \Ma^2}\RF^2+\al_4\LF{\Box\over \Ma^2}\RF^4\RT\,.
\label{caseA}
\ea
By solving (\ref{tr-cons1}) and (\ref{tr-cons2}) for the hyperbolic cosine bounce,  and imposing
\be
\Ga_0=\exp\LT\al_1\LF{\la\over \Ma^2}\RF+\al_2\LF{\la\over \Ma^2}\RF^2+\al_4\LF{\la\over \Ma^2}\RF^4\RT\,,
\ee
we can parameterize the $\al$'s as
\ba
\al_1&=& {3\over 2}\Ga_0-1\,, \\
\al_2&=&1+2\ln\Ga_0-{9\Ga_0\over 4}+{1\over 2\Ga_0}\,, \\
\al_4&=&{3\Ga_0\over 4}-{1\over 2\Ga_0}-\ln\Ga_0\,,
\ea
where we have used/chosen
\be
\mu=-6\la^2=-6\Ma^2\,.
\ee

The bounce radiation density is given by
\be
\rho_b={3\over 2}M_p^2\Ma^2(\Ga_0-1)\,.
\ee
Thus, as long as $\Ga_0>1$ the radiation density is positive and we obtain a consistent hyperbolic cosine bounce in the nonlocal model. Below we show the values of $\al$'s as a function of the $\Ga_0$ parameter for which the solution exists. After some series manipulations, we find that the coefficients $c_n$ in the action corresponding to this theory are given by
\be
c_n = -\frac{\Ma^{2(n+1)}}{6}\sum^{n+1}_{m=0}\sum_{k=0}^{n+1}\frac{\al_1^k\al_2^{\frac{1}{4}(n+1-2m+k)}\al_4^{m-k}}{\left(\frac{n+1-2m+k}{4}\right)!k!}\,,
\ee
where only the terms where the factorial is a positive integer are counted in the sums. Thus we have shown how one may also reverse-engineer nonlocal theories to generate the desired background evolution.

In figure \ref{alphas2} we provide the range of the relevant parameters as a function of $\Ga_0$ for $\la=\Ma^2$.
\begin{center}
\begin{figure}
\begin{center}
\includegraphics[width=8.0cm]{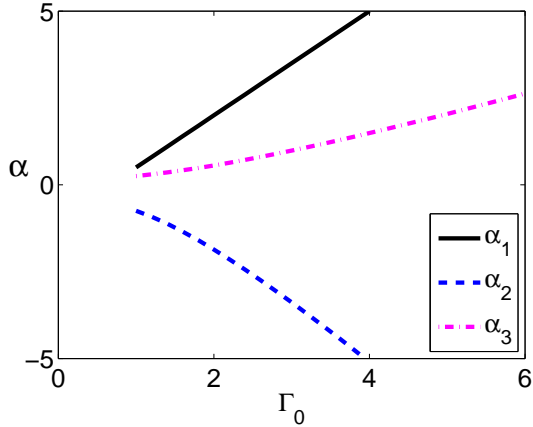}
\caption{\label{alpha2}
%Values of $\al_1$ (red), $\al_2$ (green) and $\al_4$ (brown) as a function of $\Ga_0$.}
Values of the three parameters $\al$ as a function of $\Ga_0$ as indicated in the figure.}
\end{center}
\end{figure}
\end{center}

%%%%%%%%%%%%%%%%%%%%%%%%%%%%%%%%%%%%%%%%
\section{Cosmologically viable non-local models }
%\subsection{Stability \& and Ghost free Radiation}

To summarize, we have found that given a specific higher derivative theory of the form (\ref{non-pert}), \ie given a specific $\Ga(\Box)$ function, we can find a one parameter family, labeled by the first derivative of the Hubble parameter at the bounce, of symmetric bouncing solutions. Each coefficient $h_{2n+1}$ is uniquely determined by the previous ones, and thus one may extend the solution in principle arbitrarily late in time.  However there are several important questions one needs to answer before we can start taking them as serious candidates for consistent and phenomenologically viable non-singular theories. In this section we want to discuss these issues.

The first question has to do with the radiation density at the bounce which is fixed in terms of the model parameters and initial conditions (this could be turned the other way around too), and is given by (\ref{rhonot}). Since we want to consider theories without ghosts, this radiation energy density must be positive which provides additional constraints on the parameters. The second issue has to do with
whether suitable GR limits of the solutions exist?  Finally, it is important to find out whether the bouncing solutions that we have found are generic, or are there background solutions which end up in a singularity? In this section we will try to address these questions staying within the confines of our ansatz equation.

%%%%%%%%%%%%%%%%%%%%%%%%%%%%
\subsection{$\la<0$ Case, problems recovering a GR limit}
\label{la<0}

Before we discuss the non-local theories let us revisit the much studied $R^2$ modification of gravity:
\be \label{quad}
F(R)=R+{c_0\over \Ma^2} R^2\,,
\ee
%where $\alpha=-c_0/\Ma^2$.
%Thus the ansatz fixes only the propagator and its first derivative at a single point, the pole, and seems not to be restrictive.
In Ref.\cite{Page:1987cb} it was shown that a zero-measure set of nonsingular solutions exists for this model.
Even in the absence of radiation and a cosmological constant, the $\cosh$-bounce is a solution in a closed universe \cite{Starobinsky:1980te}.
 Nonsingular solutions and attractor solutions for conformally invariant fields have been found also taken into account quadratic corrections to the gravity action \cite{anderson}, see also \cite{Rsquared}. The bounce solution that we have for $R^2$ gravity is a reiteration of these results.

This quartic higher derivative action is actually unique because it is the only case when the solutions are not over determined:
\be \label{lamu}
\la= {\Ma^2\over 6c_0} \mand \mu=-{2\La \Ma^2\over 3c_0M_p^2}\,.
\ee
This was checked in the appendix \ref{app4}.
%For the second case, our ansatz only works when $\La=0$ but we obtain precisely the same  values for $\la,\mu$ as the first case. Thus, the ansatz for both  classes of higher derivative theories is identical, and therefore so are the solutions!
%In the case of vanishing cosmological constant the expressions somewhat simplify,
%\be \label{rhop}
%\rho =  \frac{3 M_p^2}{\la}\LF H(H\la+2\ddot{H})+6H^2\dot{H}-\dot{H}^2\RF\,, \quad p = \rho/3 \,.
%\ee
One can also check that there is a large parameter space where the radiation density is positive. For instance, for the special case when the cosmological constant and $\mu$ vanishes, the constraint (\ref{rhonot}) gives us $\rho_0=-3M_p^2h_1^2\oa^4/\la$. Thus as long as $\la<0$ we can have a consistent bouncing solution. This is rather puzzling because it is also known that in a flat universe, if we allow only a positive cosmological constant and radiation density, there are no viable non-singular solutions (even for slightly more general models with an additional term due to conformal anomaly) \cite{Fischetti:1979ue}. So, does this mean that just the simple $R^2$ modification can avoid the Big Bang singularity problem after all?

%A maximum curvature follows then by demanding that $F'(R)$ is always positive. We have $R_{MAX}=\frac{1}{2\alpha}=|3\la |$. Given these conditions, we can have a bounce in curvature-squared gravity with radiation.

The answer turns out to be rather interesting. Let us try to understand the attractor behavior of the ansatz equation when $\la<0$. Let us start with a late time trial solution
\be
H(t)=h_1t\,.
\label{super}
\ee
 (\ref{ansatz}) then reads
$$-72h_1^3t^2=12\la h_1^2t^2+\cO(1)$$
Thus we have a late time super-inflationary attractor solution given by
\be \label{super2}
h_1\ra -{\la\over 6}\,.
\ee
In the expanding branch, since $h_1>0$, this means that the attractor solution only exists if $\la<0$. Our numerical studies confirm this behavior, see figure \ref{evolution}, and we will also provide a ``Dynamical System Analysis'' further corroborating it.

The upshot is that as long as $\la<0$, although these models admit nonsingular bouncing solutions, they are phenomenologically not viable because we cannot recover the GR phase but rather are stuck in the QG phase forever\footnote{However, when considered within the first order formalism, which is an alternative way to avoid some problems in higher (finite) derivative extensions to gravity,
the quadratic model of the form (\ref{quad}) does generate a bounce with a smooth transition to the GR phase \cite{pala}.}. Hence, from now on we are going to focus on non-local theories with positive $\la$. The difference to pure $R^2$ gravity is clear from Eq.(\ref{ansatz}). This is the trace equation for $R^2$-gravity, but in that case both $\lambda$ and $\mu$ are uniquely fixed by (\ref{lamu}), and one inevitably obtains negative radiation energy density for the nonsingular solutions. In a nonlocal theory of the form (\ref{action}), $c_0$ of course doesn't alone fix these coefficients, and hence we can get bouncing solutions with just positive-density sources.

%%%%%%%%%%%%%%%%%%%%

\subsection{$\la>0$ case, and the issue of singularities}

We have seen that if $\la>0$ the QG super-inflationary late time attractor no longer exists paving the path to recover GR cosmology at late times. However, we still have to worry about singularities? Having a single bounce does not guarantee that the evolution encounters no singularities in the future (or the past). Can we find if and when we have completely nonsingular evolutions? In order to answer this we will look at the two cases of positive and negative cosmological constant separately.

\subsubsection{$\La>0$, Geodesically complete inflation}
For inflationary applications we will be interested in the case when $\La>0$. The expression for the cosmological constant (\ref{tr-cons3}) then tells us that $\mu<0$. From (\ref{rhonot}) it is now clear that to have a positive $\rho_0$
$$
\la - 2c_0\mu/\Ma^2<0\,.
%\label{inequality}
$$
Since $\mu<0$ and $\la>0$, this means $c_0<0$. Also note that we can always choose $|c_0|=1$ by choosing $\Ma$ appropriately. Hence  setting $c_0=-1$, the  inequality is given by
\be
\mu<-{\la\Ma^2\over 2}\,.
\label{inequality}
\ee

Now, since our universe is filled with radiation and vacuum energy, for a positive cosmological constant the GR solution corresponds to a deSitter solution. It is obvious that one of the late time solutions is the deSitter universe:
\be
H=e^{\pm t\sqrt{\La/3M_p^2}}\,.
\ee
The important question we want to find out is how generic and stable this late time deSitter solution is.
 Both numerical and analytical studies confirm that this is indeed a stable attractor, we will see this in the next subsection \ref{sec:dynamical}. However,  are there other singular attractor solutions which may sometimes prevent the universe to reach the above deSitter background?

Consider the ansatz
\be
H(t)={p\over t}
\label{big-crunch}
\ee
Then as $t\ra 0$, the most singular contribution in the ansatz equation (\ref{ansatz}) is $\cO(1/t^4)$. If we demand that the coefficient of the $\cO(1/t^4)$ term must vanish we end with two possibilities:
\be
p=\2\mx{ or }p=1\,.
\ee
The first one is the usual GR solution in radiation dominated era. This is in fact an exact solution of (\ref{ansatz}) when $\La=\mu=0$, but crucially this solution is unstable, as we will discuss in the next subsection. The second solution however represents a Big crunch singularity. Can these singular solutions be projected out when we restrict ourselves to having positive radiation energy densities? A simple way to check this is to compute the leading singular behavior in $\rho_r$ as we approach the singularity. From (\ref{rho1}) we find
\be \label{leading_singular}
\frac{\rho_r}{M_p^2} =  -27\frac{\la+2\mu/\Ma^2}{t^4(3\la^2-\mu)} +\cO\LF{ 1\over t^2}\RF
%\frac{\rho_r}{M_p^2} =  9\frac{\la+2\mu/\Ma^2}{t^4(3\la^2-\mu)}\LF{-\frac{3}{t^4} + \frac{\la}{t^2}}\RF this includes the other singular term, but we don't need it -Tomi
\ee
(\ref{inequality}) then implies that radiation is indeed positive. Thus we cannot exclude the existence of singular solutions and they represent genuine attractors. However, as will be shown in detail below, for symmetric bounces we are considering, the singularity is never reached, and the universe evolves asymptotically towards the global GR attractor solution at low curvature. Highly asymmetric bounces however can sometimes end in a singular Big Crunch singularity, but there is a large region in the parameter space of initial conditions for which we reach the deSitter phase asymptotically at late times. This will be discussed in more details in section \ref{sec:dynamical}. Finally, one may worry about other higher order singularities ($H\propto 1/t^n$ with $n>1$), but simple counting arguments show that such singularities don't exist.

Incidentally, using (\ref{inequality}) one can also obtain a lower bound on the value of the cosmological constant. We have
\be
\La = -{\mu M_p^2\over 4\la}\geq {\Ma^2M_p^2\over 8}\,.
\ee
Note that we can adjust the cosmological constant to lower values by lowering the scale $\Ma$ (or equivalently, by keeping $\Ma$ fixed but allowing higher quadratic term $c_0$ if one doesn't choose the convention $|c_0|=1$ we used in this subsection).
%%%%%%%%%%%%%%%%

\subsubsection{$\La<0$, and cyclic universes}

\begin{center}
\begin{figure}
\begin{center}
\includegraphics[width=9.0cm]{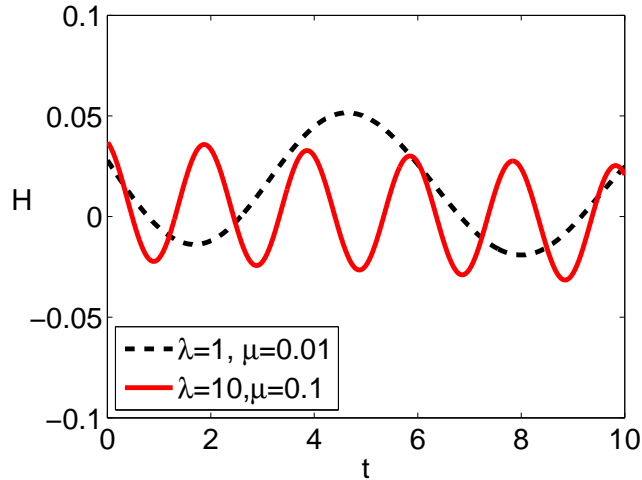}
\caption{\label{cycles}
Cyclic evolution for some parameter combinations $\al,\mu>0$. The Hubble rate $H$ and the time $t$ are in units of $\Ma$.}
\end{center}
\end{figure}
\end{center}

A rather interesting case emerges if one considers $\la,\mu>0$. In this case there are no QG super-inflating attractors, and neither do the singular solutions exist for positive  radiation energy densities, see Eq.(\ref{leading_singular}). What we obtain instead are cyclic universes, where QG effects trigger a bounce, while the universe turns around once the radiation energy density dilutes and the negative cosmological constant cancels it. In figure \ref{cycles} we show numerical solutions of such an evolution.

What is especially nice is that there is a special  class of non-singular actions with $\la,\mu>0$, and $\mu<3\la^2$ which does not admit any singular solution. This is because (\ref{leading_singular}) tells us that singular solutions can only exist for negative radiation densities. Conversely, if we have standard radiation, such solutions do not exist.

Phenomenologically, if we just have a negative cosmological constant, it is not interesting because we end up in an eternal cyclic scenario where the period is too short for structure formation. However, recently similar scenarios have been discussed where once one takes into account of interaction between different species and entropy production, viable cosmological scenarios emerge~\cite{BM}.
%%%%%%%%%%%%%%%%

\subsection{Dynamical system analysis}
\label{sec:dynamical}
The asymptotic state of the system can be studied by identifying the attractors in the phase space.
In terms of the dimensionless variables
\be
\epsilon \equiv -\frac{\dot{H}}{H^2}\,, \quad \Omega_R \equiv \frac{\rho_R}{3M_p^2 H^2}\,, \quad \Sigma \equiv \frac{1}{\xi H^2}\,,
\ee
where we have defined the auxiliary quantity
\be
\xi = 9\frac{\la-2c_0\mu/\Ma^2}{3\la^2-\mu}\,,
\ee
the ansatz, Friedmann and continuity equations can be written as
\ba
\Sigma' & = & 2\en\Sigma\,, \\
\Omega_R' & = & 2(\en-2)\Omega_R\,, \\
\en'  & = & \frac{3}{2}(\en-2)\en + \frac{1}{2}(\xi\la - 3\Omega_R + \frac{1}{12}\xi^2\mu\Sigma)\Sigma \,, \\
\en'' & = &  \left[7 \en' + 6 (2 - \en) \en\right](\en-1) + (2 - \en)\xi\la\Sigma + \frac{1}{6} \xi^2 \mu \Sigma^2 \,.
\ea
The system has the following fixed points:
\begin{itemize}
\item $A$: The radiation dominated solution $\epsilon=2$, $\Sigma=0$. The linearized matrix for perturbations about this fixed point
has two positive eigenvalues, meaning the solution is unstable.
\item $B$: An inflating attractor $\Sigma=\en=\Omega_R=0$, which is stable.
\item $C$: The deSitter solution $\Sigma=-12\la/(\xi\mu)$, $\en=\Omega_R=0$. It is an attractor when $\mu<0$, otherwise a saddle point.
\end{itemize}

To understand the meaning of these fixed points, it is useful to consider the expansion-normalized energy contributions
\be
\tilde{\Omega}_R=\frac{\rho_R}{|\xi\la|M_p^2 H^2}\,, \quad \Omega_\La = -\frac{\mu}{12\la H^2}\,, \quad Q_1 = -\frac{1}{\la}\dot{H}(6+\en)\,, \quad Q_2 = -\frac{2}{\la}\frac{\ddot{H}}{H}\,.
\ee
The first two are the relative contributions from radiation and the cosmological constant, respectively, and the latter two represent the non-local corrections to
gravity\footnote{Despite this intuitive form, these turn out to be less suitable for dynamical system analysis, since the phase space spanned by these variables contains only the fixed point $C$.}.
In terms of these, the Friedmann equation is
\be
1=sgn(\xi\la)\tilde{\Omega}_R+\Omega_\La+Q_1+Q_2\,.
\ee
The sign appears in front of $\tilde{\Omega}_R$ since its contribution can be negative if $\xi\la<0$, though we always consider positive physical radiation energy density.
The unstable fixed point $A$ corresponds to $\tilde{\Omega}_R=1$, the attractor $B$ to $Q_1=1$ and the fixed point $C$ to $\Omega_\La=1$. It is then clear that $B$ is the solution (\ref{super},\ref{super2}) we derived previously. Since we are interested in the post-bounce evolution, we consider the expanding branch, where the fixed point $B$ only exists for negative  $\la$. Then, the universe will be super-inflating, but approaching $\en \rightarrow 0$ instead of ending in a Big Rip. This possibility would be interesting in the dark energy context,
%\footnote{Nonlocal cosmology with possible phantom crossings were considered in e.g. \cite{nonlocalities}.},
but in the present study we do not regard evolutions ending in $B$ as  viable trajectories, since the phantom expansion is driven by $Q_1$ and we want to recover pure GR.

\begin{center}
\begin{figure}
\includegraphics[width=8.0cm]{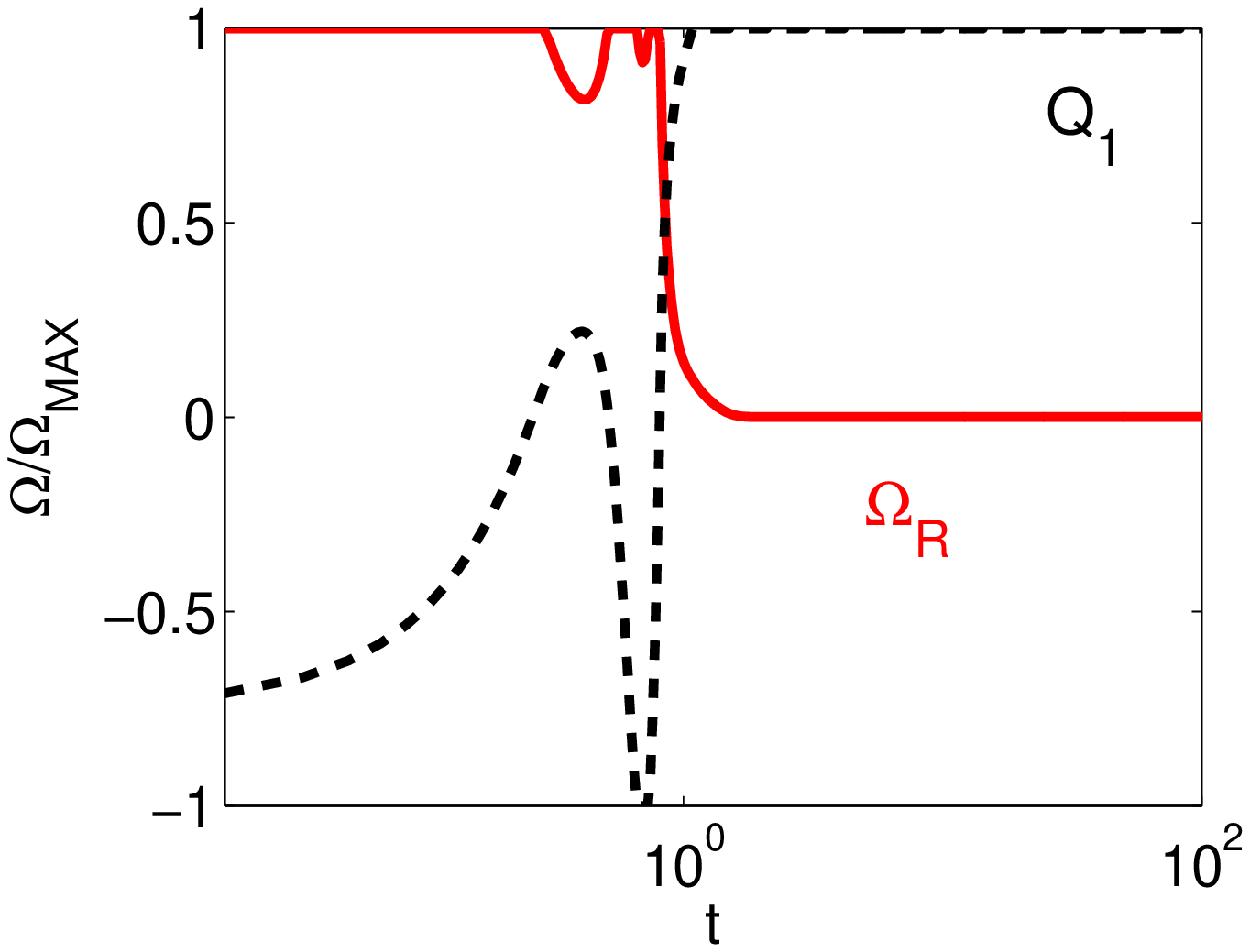}
\includegraphics[width=8.0cm]{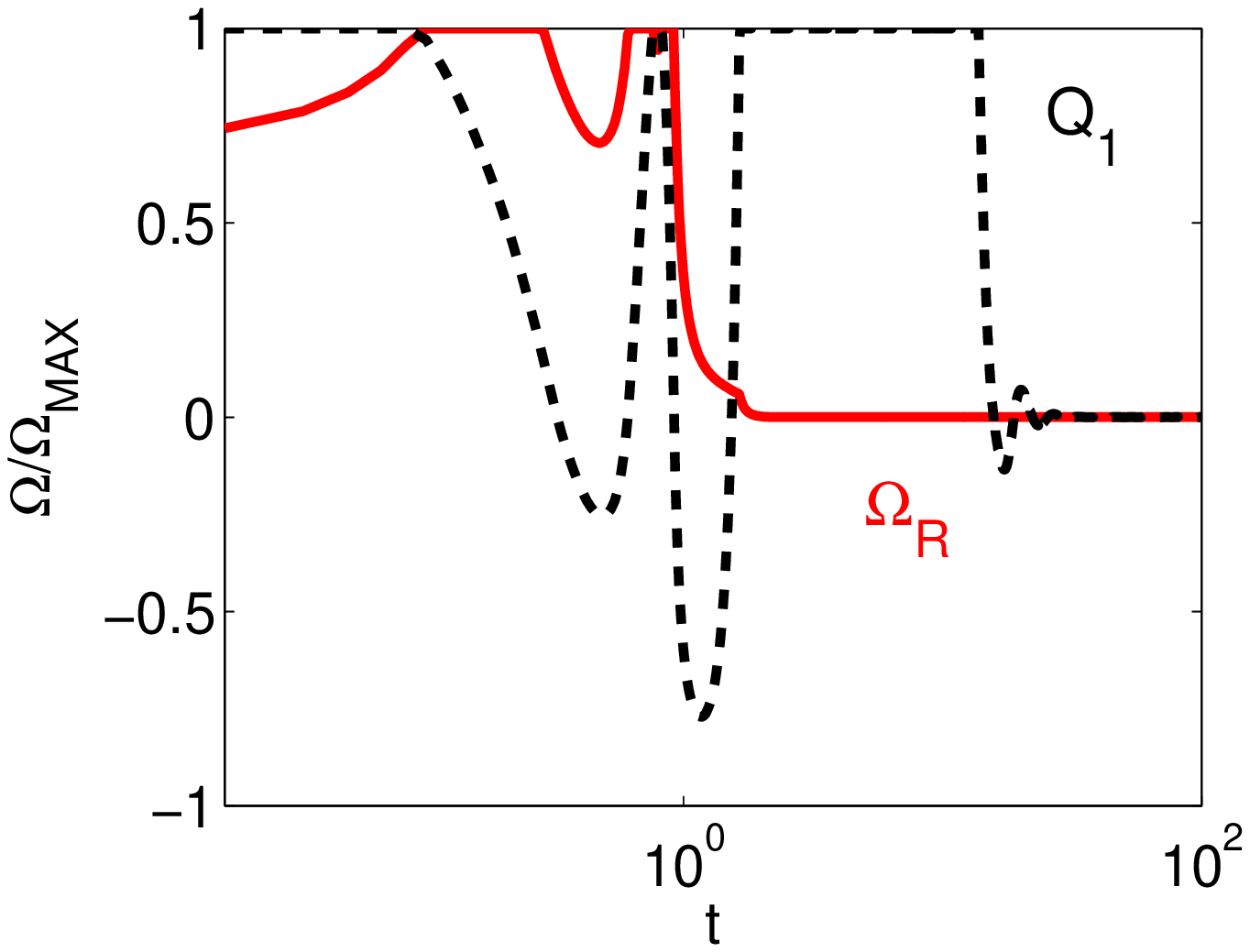}
\caption{\label{evolution}
Dynamics at the bounce for two models for normalized relative energy contributions, $Q_1$ being the nonlocal and $\Omega_R$ the radiation contribution. Left panel: evolution in the $\la<0$, $\mu<0$ case: the super-inflating solution $B$ is reached after the bounce. Right panel: the same model but with flipped sign of $\la$. Now the de Sitter $C$ attractor will be reached, and both $Q_1$ and $\tilde{\Omega}_R$ vanish asymptotically. }
\end{figure}
\end{center}

\begin{center}
\begin{figure}
\includegraphics[width=8.0cm]{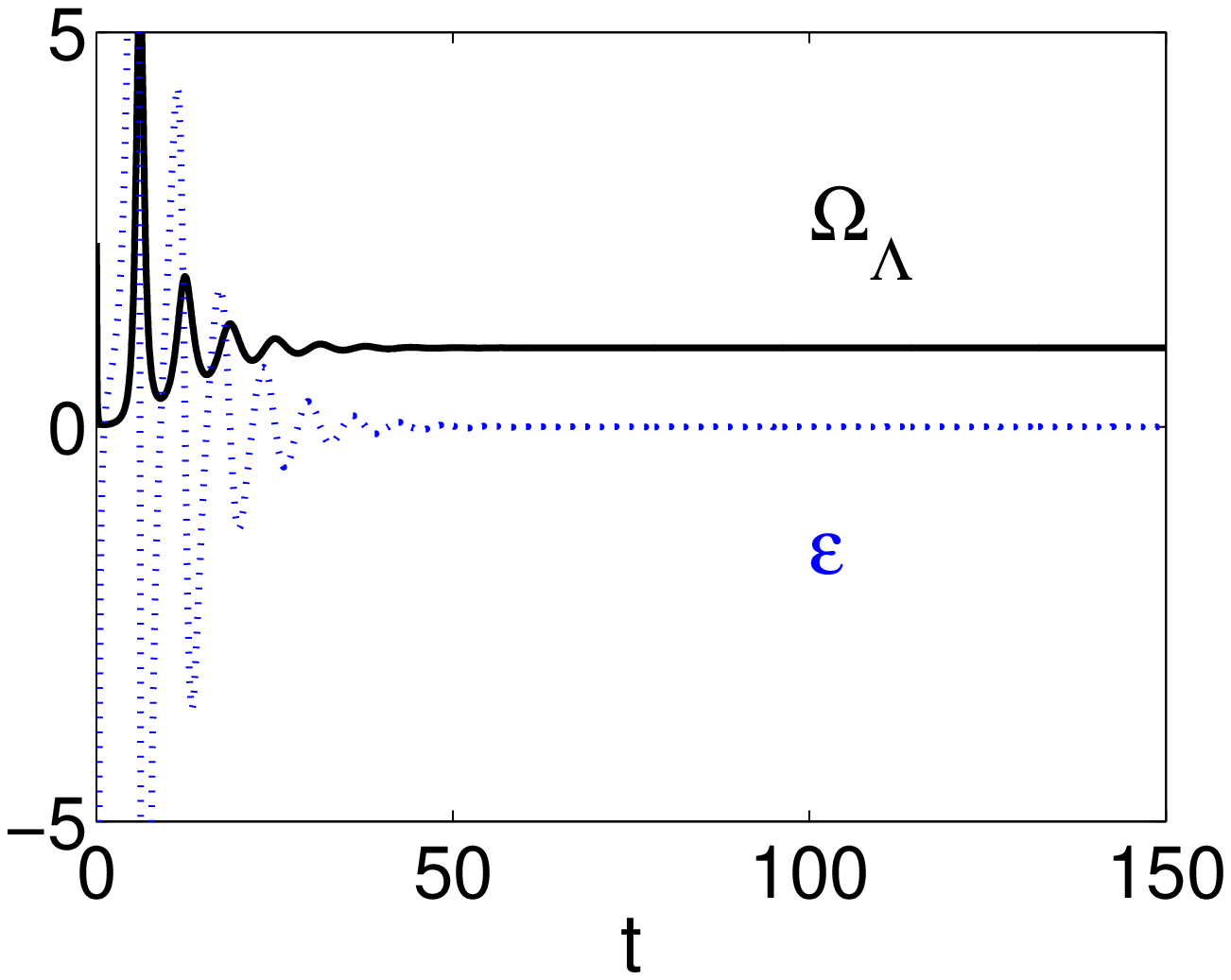}
\includegraphics[width=8.0cm]{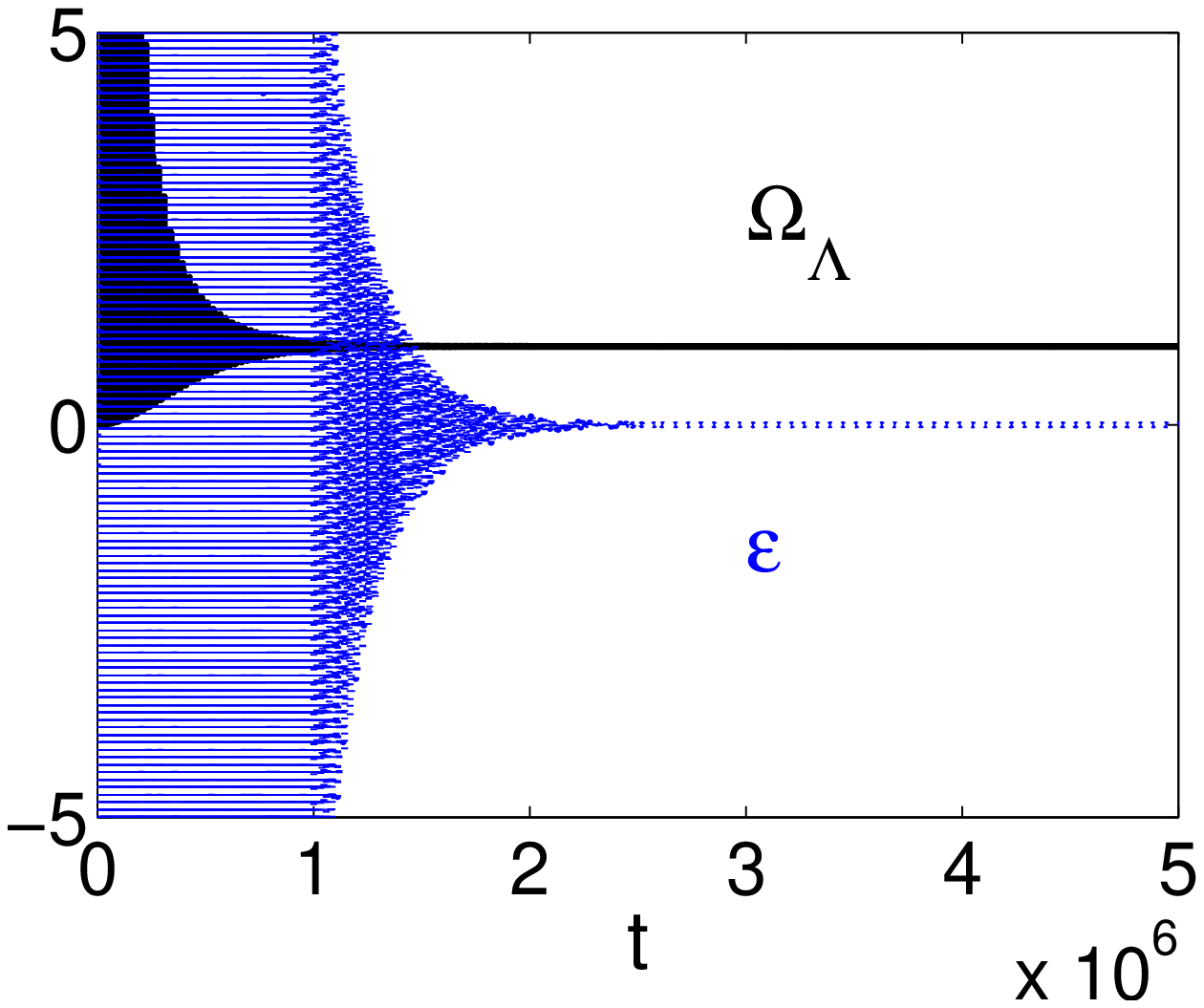}
\caption{\label{muplot}
Convergence to the deSitter attractor $C$ when $\la=1$. In the left panel $\mu=-0.1$, and in the right panel $\mu=-e^{-10}$. The oscillations take longer for smaller $\mu$, but the GR attractor $C$ is reached eventually.}
\end{figure}
\end{center}

\begin{figure}
\begin{center}
\includegraphics[width=9.0cm]{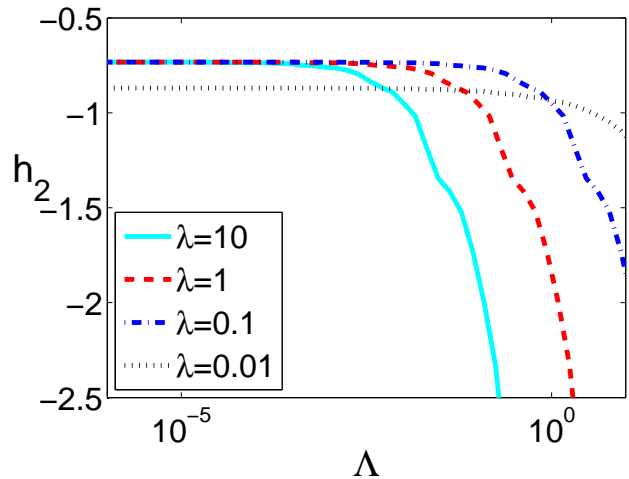}
\caption{\label{h2plot}
Lower limit on the initial perturbation $h_2$ to converge to the GR attractor $C$ when units are such that $h_1=1$ at the bounce. Above the lines, we reach the deSitter fixed point $C$. Below the lines, with large negative $\ddot{H}$ (presumably an unnatural boundary condition at the bounce), we are driven to the big crunch. This shows that the GR attractor $C$ is stable to small perturbations.}
\end{center}
\end{figure}

Thus, to underpin our reasoning in the preceding subsections, we have shown that we need a positive $\la$ to avoid the attractor $B$ and a negative $\mu$ to be attracted towards the GR evolution $C$ where the nonlocal terms $Q_1$ and $Q_2$ both vanish. Example evolutions of bounces evolving to $B$ and to $C$ are shown in Fig. \ref{evolution}.
Numerical experiments confirm that the GR phase $C$ will be generically approached after the turnover, given just the correct sign choices. The cosmological constant need not to be large for this, but can be adjusted to the desired scale. Evolutions for both large and small $\Lambda$-terms are shown in Fig. \ref{muplot}. Moreover, the convergence to the attractor is insensitive to small perturbations in the initial conditions. Since $H$ vanishes at the bounce and we can normalize $\dot{H}$ to unity in units of $\Ma$, the most general initial conditions are given by varying $\ddot{H}$ at the bounce, or equivalently, considering nonzero $h_2$. Only if $h_2$ is large and negative, we fail to recover the GR regime, as illustrated in Fig. \ref{h2plot} for several choices of $\la$.


We remind that we have only exhausted the possibilities within ansatz we have made. In reality the dynamics will change as matter becomes non-relativistic and dust sources render our system based on the ansatz invalid. Furthermore, even when the energy sources consist only of radiation and $\Lambda$, there can be solutions which do not obey equations (\ref{tr-cons1}-\ref{tr-cons3}), i.e. not each of the contributions to the trace equation vanishes identically. Our stability analysis here concerns only the reduced phase space where the trajectories obey these conditions and the ansatz (\ref{ansatz}). In principle this class of trajectories need not be stable among the space of all solutions. At a more fundamental level, ghost-free condition of section (\ref{ghostfree}) is only about the Minkowski space, which may not be enough to guarantee that ghosts wouldn't appear at a nonlinear level in strongly curved backgrounds. In a complete analysis, one would have to include perturbations in the metric as well and allow inhomogeneous modes. To fully address these problems, one has to device other techniques to solve the full non-local ES differential equations. This is beyond the scope of our present paper.
%In fact the two-parameter family satisfying the ansatz could be unstable in the space of all solutions.
%%%%%%%%%%%%%%%%%%%%%%%

%%%%%%%%%%%%%%%%%%%%%%%%%%%%%%%%%%%%%%%%%%%%%%%%%%
\section{Perturbations}

In this section, we show how the non-singular bouncing solutions satisfying our ansatz can be used to study the evolution of large-scale perturbations by either series methods or by numerically solving differential equations. %Perturbations in non-local cosmology were discussed also in \cite{perturbations}.

\subsection{Tracking perturbations across bounce}
We  want to investigate  how  we can track the evolution of small  perturbations around a specific bouncing background solution. This is obviously a hard problem and we fall back on our intuition that was developed by dealing with scalar fields. In the scalar field case perturbations around a background satisfied the same local differential equation as the background. Since we are looking at small perturbations around the background solution which satisfies a ``local field equation'', it seems likely that the perturbed solutions also satisfy the same ``local field equation'', by virtue of which they will also solve the full non-local equations. This is where the covariant nature of our ansatz comes into play. This strategy was successful in the scalar field case and for our gravity theory, in analogy,  all we need to do is to perturb our ansatz (\ref{ansatz}). In this approach,  at worst, we might be throwing away some modes of perturbations (which solve the full ES equations (\ref{e-s}), but not the ansatz equation (\ref{ansatz})), but at least we should be able to successfully track the ones which satisfy the ansatz equation.  A further drastic simplification arises if we assume that there are no more than $\sim 70$ e-foldings of inflation after the bounce. In this case, the physical wavelengths of the perturbations, that we see today in the CMB sky, at the bounce point are still too large as compared to the Planck length. In other words,  the perturbations behave as super-Hubble modes during the bounce, the time scale of bounce is too short as compared to the physical wavelength. We can therefore ignore any spatial dependence of the perturbations and just look at the fluctuations of the zero-mode or the scale factor. (Note, if the fluctuations in the metric just depend on time, one can always perform a coordinate transformation such that the only independent function is the scale factor.). Thus, just by looking at time dependent fluctuations of the ansatz equation we can unambiguously track the perturbations across the bounce, and this formalism will be explained in details in~\cite{abhay}.
%%%%%%%%%%%%%%%%%%%%%
%\subsection{Super-Hubble fluctuations}

Now fluctuations of the scale factor, $\da(t)$, can be encoded as follows:
\be
ds^2=-dt^2+a^2(t)dx^2\mx{ with }a(t)=\bar{a}(t)[1+\da(t)]\,,
\ee
so that
\be
H(t)=\bar{H}(t)+\dot{\da}\,.
\ee
The linearized expressions for $R$ and $\Box R$ is given by
\ba
R&=&\bar{R}+6(\ddot{\da}+4\bar{H}\dot{\da})\equiv \bar{R}+\da R\,, \nonumber\\
\Box R&=&\bar{\Box}\bar{R}-(\ddot{\da R}+3\bar{H}\dot{\da R}+3\dot{\da}\dot{\bar{R}})\,.
\ea
Thus the perturbation equation reads
\be
\ddot{\da R}+3\bar{H}\dot{\da R}+3\dot{\da}\dot{\bar{R}}=-\la\da R\,.
\label{pert}
\ee
This is a linear fourth order differential equation and in principle can be solved. In practice, one has to again employ power-series methods to solve for $\da(t)$. However, it is clear that a trivial solution of the above equation is $\dot{\da}=0$, or $\da$ is a constant. This is analogous to the statement in GR that the super-Hubble perturbations remain constant. However, a crucial thing one has to check is whether any of the above modes are spurious or not. In other words, one has to substitute $\da(t)$ obtained by solving (\ref{pert}) in the $\w{G}_{00}$ equation and check that it is also satisfied, just as we did for the background.

Let us then expand $\da$ as
\be
\delta(t) = \da_1(\Ma t) + \da_2(\Ma t)^2 + \da_3(\Ma t)^3 + \cdots\
\ee
From zeroth to the third order, the coefficients in the expansion of the perturbed ansatz equation (\ref{pert}) as a power series in $t$ then give
\ba
0 & = & \da_2 \la  + 4 (2\da_2h_1) \Ma^2\,, \\
0 & = & 7 \da_1 \mu  + 18 (\da_1 h_1-2 \da_3)\la \Ma^2 - 12 (60 \da_5 + h_1 (45 \da_3 - 2 \da_1 h_1)) \Ma^4\,, \\
0 & = &  11 \da_2 \mu  + 6 (7\da_2h_1-6 \da_4) \la \Ma^2 - 24 \left[45 \da_6 + h_1 (33 d_4 - 2 \da_2 h_1)\right] w^4\,, \\
0 & = & -\frac{1}{2} \da_1 \la \mu  + \frac{1}{3} \left[
 156 \da_3 \mu - \da_1 h_1 (9 \la^2 + 29 \mu)\right] \Ma^2 -
   6 \left[ 20 \da_5 + h_1 (7\da_1h_1-40 \da_3)\right]
    \la \Ma^4 \nonumber \\ & - &
   8 \left[ 630 \da_7 + h_1 (435 \da_5 + 8 h_1 (\da_1 h_1-6 \da_3))\right]\,.
\ea
where we have used the solutions (\ref{hh_3}-\ref{hh_7}) for the background quantities in terms of $h_1$. From the structure of the expansion
we see that once $\da_2$ is given, each $\da_{2n}$ where $n>1$ is determined uniquely. Since $\da_{2n}$ appears linearly in the coefficient
of the $t^{2(n-2)}$ term, the solutions are guaranteed to be real. Similarly, given $\da_1$ and $\da_3$, each $\da_{2n+1}$ where $n>0$ is determined
by the constraint from the $t^{2n-3}$ coefficient. To fix $\da_2$, we can use the Friedmann constraint at the zeroth order. This gives
\be \label{p_fri}
3 (\la - 2 c_0 \mu/\Ma^2) \left[\da_0 \mu - 12 h_1 (\da_2 + \da_0 h_1) \Ma^4\right]=0\,.
\ee
This allows us to solve $\da_2$ in terms of $\da_0$
(barring the special case $\Ma^2\la=2c_0\mu$, when both $\da_0$ and $\da_2$ will be left as free parameters).
It follows that all $\da_{2n}$  will be proportional to $\da_0$, which may be regarded as an initial condition.
Up to the seventh order, the full solution is then explicitly
\ba
\da_2 & = & \da_0 \LF-h_1 + \frac{\mu}{12 h_1 \Ma^4}\RF\,,  \label{da_2} \\
\da_4 & = & \da_0 \frac{(\la + 8 h_1 \Ma^2) (-\mu + 12 h_1^2 \Ma^4)}{144 h_1 \Ma^6}\,, \\
\da_5 & = & \frac{7 \da_1 \mu + 18 (\da_1 h_1-2 \da_3) \la \Ma^2 + 12 h_1 (2\da_1h_1-45 \da_3) \Ma^4}{720 \Ma^4}\,, \\
\da_6 & = & \da_0\frac{(\mu - 12 h_1^2 \Ma^4) (3 \la^2 + 11 \mu + 132 h_1 \la \Ma^2 + 576 h_1^2 \Ma^4)}{12960 h_1 \Ma^8}\,, \\
\da_7 & = & \frac{1}{30240\Ma^6}\Big[-10 \da_1 \la \mu + 3 \left[12 (\da_3 - \da_1 h_1) \la^2 + (104 \da_3 - 87 \da_1 h_1) \mu\right] \Ma^2 \nonumber \\
& + & 42 h_1 (72 \da_3 - 19 \da_1 h_1) \la \Ma^4 + 36 h_1^2 (499 \da_3 - 30 \da_1 h_1) \Ma^6\Big] \label{da_7} \,.
\ea
This could be continued to track the evolution further from the bounce at $t=0$.
Only the constant part of the Friedmann constraint, (\ref{p_fri}) contained nontrivial information. The solution (\ref{da_2}-\ref{da_7}) satisfies
the time-dependent part of the Friedmann constraint identically. We also note that as for the background, the details of the higher order derivatives,
i.e. $c_n$ where $n>0$, are not important for the perturbations.
This is because the superhorizon modes have the same symmetries as the background.
Inhomogeneous perturbations could in principle be sensitive to the higher order terms in the action.

To summarize, once the background evolution is known, by fixing three initial conditions the perturbations can be determined order by order by the method above. Alternatively,
one can integrate the differential equation
\ba \label{nonl_pert}
\da^{(4)} & + & 7\left(H+\dot{\da}\right)\da^{(3)}+4\left(\ddot{\delta}+6H\dot{\delta}+3\dot{\da}^2+2\dot{H}+2H^2 + \la\right)\ddot{\delta} +
 2 (6 \dot{H} + 6\ddot{\delta}  + \la)\dot{\delta}^2  \nonumber \\  & + &
  \left(7 \ddot{H} + 4 H (6 \dot{H} + \la)\right)\dot{\delta} = 0\,.
  %\delta^{(4)} + 7H \delta^{(3)}  + (8 \dot{H} + 12H^2 + \lambda + 4 \ddot{\delta})\ddot{\delta}  + 2 (6 \dot{H} + 6\ddot{\delta}  + \la)\dot{\delta}^2   +
%\left(7 \ddot{H} + 7 \delta^{(3)} + 4 H (6 \dot{H} + 6 \ddot{\delta}  + \la)\right)\dot{\delta} = 0\,.
\ea
where we have omitted the bars from background quantities. In the linear approximation this simplifies to
\be \label{lin_pert}
\da^{(4)}  +  7H\da^{(3)} + (8\dot{H}+12H^2+\la)\ddot{\da} + \left(7\ddot{H}+4H(6\dot{H}+\la)\right)\dot{\da} = 0\,.
\ee
Below we provide analytical and numerical solutions to these equations.

%%%%%%%%%%%%%%%%%%%%%%%
\subsection{Example of hyperbolic cosine solution}
\begin{figure}
\begin{center}
\includegraphics[width=9.0cm]{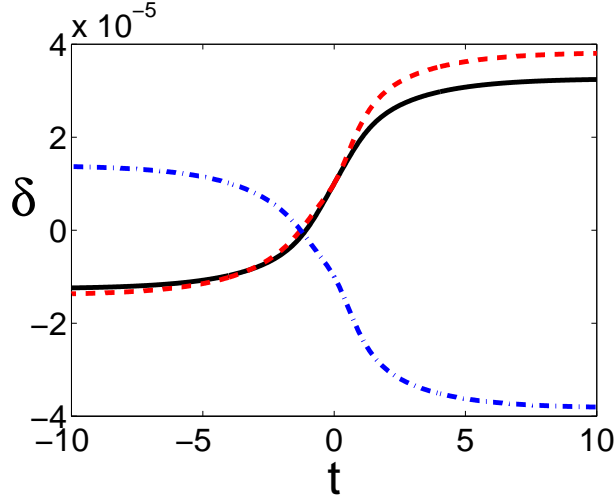}
\caption{\label{deltapic}
Numerical solutions of the perturbation evolution in the $\cosh$-bounce for different initial conditions at $t_i=0$. Units are such that $\Ma=1$.
%Solid lines are solutions to the full nonlinear equation (\ref{nonl_pert}), dashed lines to the linear approximation (\ref{coshpert}).
Solid (black) line: $\delta_i=\dot{\delta}_i=1.0e^{-5}$. Dashed (red) lines: $\delta_i=\dot{\delta}_i=\ddot{\delta}_i=\delta^{(3)}_i=1.0e^{-5}$. We also plot a decaying mode with dash-dotted (blue)
line: $\delta_i=\dot{\delta}_i=-1.0e^{-5}$.
Note that the linear perturbation $\delta$ can always be rescaled as the normalization does not affect the evolution.
%The results show the existence of a linear growing mode in the contracting phase and
The results confirm that the superhorizon perturbations freeze in the expanding phase.}
\end{center}
\end{figure}
A more complete analysis of perturbations will be presented elsewhere, but here we look at a special case to understand how the perturbations evolve. Let us go back to the hyperbolic cosine bounce, so that (\ref{lin_pert}) reads
\ba \label{coshpert}
\da^{(4)}  &+&  7\sqrt{\la\over 2}\tanh\LF\sqrt{\la\over 2}t\RF\da^{(3)} + \nonumber
\LT4\la \sech^2\LF \sqrt{\la\over 2}t\RF+6\la \tanh^2\LF\sqrt{\la\over 2}t\RF+\la\RT\ddot{\da}\\
 &+& \Big[{7\la^{3/2}\over \stwo}\tanh\LF\sqrt{\la\over 2}t\RF \mt{sech}^2\LF\sqrt{\la\over 2}t\RF \nonumber \\
& + & 4\sqrt{\la\over 2}\tanh\LF\sqrt{\la\over 2}t\RF\LF3\la \mt{sech}^2\LF\sqrt{\la\over 2}t\RF+\la\RF\Big]\dot{\da} = 0\,.
\ea
The above equation can be solved numerically. We also show numerical examples of the perturbation evolution in figure \ref{deltapic} for three choices of initial conditions. As the perturbations are small, the solutions of both the linear equation (\ref{lin_pert}) and nonlinear equation (\ref{nonl_pert}) agree (we have checked that in the case of large perturbations, the effect of nonlinearities is to stabilize the growth). We see that there exists a linear growing mode in the contracting phase. Consequently, the amplitude of the perturbations is enhanced before they  freeze in the expanding phase to a constant value, precisely what one would like for the inflationary mechanism to work.

One can actually understand this feature analytically as well. At late times, the equation simplifies considerably:
\be
\da^{(4)}  +  7\sqrt{\la\over 2}\da^{(3)} + 7\la\ddot{\da} + \left(4\sqrt{\la^3\over 2}\right)\dot{\da} = 0.
\ee
The solutions are of the form $\da=\exp(\nu t)$ with
\be
\nu^{4}  +  7\sqrt{\la\over 2}\nu^{3} + 7\la\nu^2 + \left(4\sqrt{\la^3\over 2}\right)\nu = 0.
\ee
The solutions are
\be
{\nu\over \sqrt{2\la}}=\{0,-1/2,-2,-1\}
\ee
Thus all the solutions are decaying except the first one which is the usual constant mode, and the standard inflationary mechanism can operate successfully.

%%%%%%%%%%%%%%

\section{Conclusions}

In this work we have shown how ghost and asymptotically free non-perturbative effective actions of
gravity motivated from string theory are able to realize  non-singular bouncing cosmologies. We studied a class of solutions
with two parameters, which solves the non-local and non-linear equations of motion for gravity in presence of a cosmological constant and radiation energy density. In particular, a specific class of non-singular bouncing solutions is given by a hyperbolic cosine scale factor of the universe.

We studied cosmological scenarios with both positive and negative cosmological constants. In the former
case we have found that an inflationary (deflationary) solution arises as a late time attractor  after (before)
 the bounce, the non-local contributions to  gravity vanishes, and the space-time asymptotes towards the GR deSitter limit. This result is very relevant for inflationary cosmology as now time-like and null-like trajectories can be made {\it geodesically complete} within the inflationary framework.
Without the non-singular bounce the trajectories in the past light cone would always be finite, which would render inflation geodesically past incomplete. Furthermore, the Bunch-Davies vacuum would be pathological in the infrared.
%We also argued that in order to get an inflationary attractor the cosmological constant is bounded from below, i.e. $\Lambda \geq (M_{\ast}^2M_p^2)/8$, where $M_*$ is the scale at which the higher derivative operators in the action become important.

When the cosmological constant becomes negative an interesting stable solution arises which
leads to a cyclic universe in presence of radiation. Although just with radiation a cyclic universe does not lead to
any phenomenologically viable  cosmology,   exciting cosmology can arise if we add a non-relativistic matter component and interactions between radiation and matter. As shown in \cite{BM}, such an interaction would lead to a cyclic inflationary scenario with distinct signatures for the cosmic microwave background perturbations. Also, our non-local model provides a concrete setting to study how perturbations are transmitted through a nonsingular bounce which is important for cyclic and bouncing models in general. Finally, the non-local actions with $\la>0,\La<0$ and $\mu<3\la^2$, constitutes an especially nice class of actions because they don't seem to contain any singular solutions (at least within the confines of our ansatz), which is not the case for actions with $\la,\La>0$. An important task for the future is to determine whether these non-local actions admit any singular solutions at all or not.

In order to make sure that both inflationary and cyclic attractors are stable, and there exists no
instabilities during the bounce, we studied the long wavelength, super-Hubble perturbations. We found that all the modes of
the long wavelength perturbations are decaying except the one which is frozen. The presence of a constant mode is a
good news for seeding density perturbations for the cosmic microwave background radiation during, before or after the bounce. The perturbation analysis suggests that there are no pathologies at least on large scales which would affect the non-singular bounce  solution. Studying the sub-Hubble perturbations is more difficult in higher derivative theories of gravity. However, this is necessary to consider the ghost problem in the bouncing backgrounds, whereas we have yet only shown that the theories are not plagued by this problem in Minkowski backgrounds. We hope to revisit this issue in a separate publication.

To summarize, we have found  non-singular bouncing solutions for the scale factor of the universe in a special class of nonlocal gravitational models which are ghost and asymptotically free in the ultra violet, but nevertheless recover the GR limit at large distance and time scales. We also made preliminary progress to study perturbations around the bouncing background and in particular demonstrated that the bouncing solutions are late time attractors and perturbatively stable.

%%%%%%%%%%%%%%%%%%%%%%%%%%%%%%%%%%%%%%%%%

\section*{Acknowledgments}

The authors are thankful to Robert Brandenberger, Gianluca Calcagni, Alex Kusenko, Philip Stephens and Sergey Vernov for helpful discussions.
A.M. acknowledges partial support from the EU Net-work UniverseNet
(MRTN-CT-2006-035863). T.K. is supported by the FOM and the Academy of Finland.

%%%%%%%%%%%%%%%%%%%%%%%%%%%%%%%%%%%%%%%%

\appendix

\section{Condition for ''ghost-freeness''}
\label{ghost_app}

In section \ref{ghostfree}, we explained how the condition for ghost-freeness arises by looking at the field equations and identifying the poles in Eq.(\ref{phieq}). An equivalent, and perhaps a more rigorous, way to see this is to work in terms of the action involving the two auxiliary scalar fields $\phi$ and $\psi$ directly. On a Minkowskian background the  momentum space action looks like
\ba
S \, &=& \, {M_p^2\over 2}\int d^4k  \Psi_k\cK(-k^2)\Psi_k \, ,
\ea
where we have defined
\be
\Psi_k=\LF
\begin{array}{c}
\phi_k\\
\psi_k/\Ma^2
\end{array}
\RF
\mx{ and }
\cK(-k^2)=\LF
\begin{array}{cc}
{3\over 2}(-k^2) & -\2\Ma^2\\
-\2\Ma^2& \Ma^2\sum_{n = 0}^{\infty} c_n \LF{-k^2 \over \Ma^2}\RF^n\,.
\end{array}
\RF
\ee

Now, as long as the determinant of the kinetic operator has at most a single root we can be assured that there are no ghosts in the theory. One can see this in the following way. One can perform a linear transformation to diagonalize the kinetic operator so that it would be of the form
$$
\cK'=\LF
\begin{array}{cc}
\Ga_1(-k^2) & 0\\
0& \Ga_2(-k^2)
\end{array}
\RF\,.
$$
It is now clear that the number of perturbative physical degrees of freedom simply corresponds to the number of zeroes of the product $\Ga_1(-k^2)\Ga_2(-k^2)$ which is nothing but the determinant of the kinetic operator. This in turn means that we can simply count the zeroes of the original kinetic operator $\cK(-k^2)$ to determine the perturbative states. As long as we have a single or no zeroes, we are perfectly safe from any problems with ghosts. For more that two zeroes there is always a ghost while the case with two zeroes may or may not have a problem depending on the sign of the residues at the poles. We do not explore this possibility here.

Now, it is easy to check that
\be
\mt{det}(\cK)={\Ma^4\over 4}\Ga(-k^2)
\ee
and thus indeed we obtain the same condition for the action to ghost free as deduced from \ref{phieq}. It is also straightforward to show that the same condition emerges if one integrates out the field $\phi$ and considers the propagator for the field $\psi$. We then obtain the equation of motion
\be
\Box^{-1}\Ga(\Box)\psi=0\,.
\ee
In the special case $k=0$ it is not legitimate to integrate out $\phi$, and when $\sum_{n = 0}^{\infty} c_n \LF{-k^2 \over \Ma^2}\RF^n=0$ one cannot integrate out $\psi$. However, as we saw above also in these special cases the same condition for $\Gamma$ determines the nature of the propagating degrees of freedom.

\section{Series expansions}
\label{series}

For the matter content we assumed a $\Lambda$-term and conformal matter, and for the metric the ansatz (\ref{ansatz})
\be \label{a_ans}
\Box R=\la R + \mu\,.
\ee
Then the Friedmann equation (\ref{rho1}) was derived,
\ba
\frac{\rho}{M_p^2} & = & 9\frac{\la-2c_0\mu/\Ma^2}{3\la^2-\mu}\left[ H\left(H\la+2\ddot{H}\right)+6H^2\dot{H}-\dot{H}^2\right] +
\frac{1-6c_0\la/\Ma^2}{12\la^3-4\la\mu}\mu^2\,. \label{a_rho}
\ea
To look for bouncing solutions, the series expansion (\ref{H-expansion}) for $H$ was made:
\be
H=\Ma\sum_{n=0}^{\infty}h_{2n+1}(\Ma t)^{2n+1}\,.
\label{a_exp}
\ee
It follows that the scale factor behaves as
\ba
\frac{a(t)}{a(0)} & = & 1 + \frac{1}{2}(h_1 \Ma t)^2 + \frac{1}{8} (h_1^2 + 2 h_3)(\Ma t)^4 +
 \frac{1}{48} (h_1^3 + 6 h_1 h_3 + 8 h_5)(\Ma t)^6 \nonumber \\ & + &
 \frac{1}{384} (h_1^4 + 12 h_1^2 h_3 + 12 h_3^2 + 32 h_1 h_5 + 48 h_7)(\Ma t)^8 +  \mathcal{O}((\Ma t)^{10})\,.
\ea
and in the following we set $a(0)=1$. The energy density is then expanded as
\ba
\rho(t) = -\frac{\mu M_p^2}{4\la} & + & \rho_0\Big[1  -  2h_1(\Ma t)^2 + \LF 2h_1^2 - h_3\RF (\Ma t)^4 -
\frac{2}{3}\LF 2 h_1^3-3h_1h_3+h_5\RF (\Ma t)^6 \nonumber \\ \label{a_mat}
& + &  \frac{1}{6}\LF 4h_1^4-12h_1^2h_3+3h_3^2+8h_1h_5-3h_7\RF (\Ma t)^8 + \mathcal{O}((\Ma t)^{10}) \Big]\,.
\ea
The first term is the cosmological constant given by the constraint (\ref{tr-cons1}), and the rest of the terms describe the radiation density.

\subsection{Solutions for the Hubble parameter}
\label{app1}

In the following we assume $3\la^2 \neq \mu$.
To solve for the coefficients $h_{2n+1}$ we use the expansions for the matter (\ref{a_mat}) and for the Hubble parameter (\ref{a_exp}) in the two equations (\ref{a_ans}) and (\ref{a_rho}).
Order by order, the ansatz equation (\ref{a_ans}) yields the following constraints:
\ba
0 & = & \mu + 6h_1\la \Ma^2 + 12(2h_1^2+36 h_3)\Ma^2\,, \label{a_re1} \\
%0 & = & \mu + 24 h_1^2 \Ma^4 + 36 h_3 \Ma^4+6 h_1 \la \Ma^2  \,, \label{a_re1}\\
0 & = & (2 h_1^2 + 3 h_3) \la  + 6 (2 h_1^3 + 11 h_1 h_3 + 10 h_5) \Ma^2 \,, \label{a_re2}\\
0 & = & (4 h_1 h_3 + 5 h_5) \la + 6 (10 h_1^2 h_3 + 13 h_3^2 + 30 h_1 h_5 + 35 h_7) \Ma^2 \label{a_re3} \,.
\ea
Leaving $\Ma$ and $h_1$ as free parameters, as they correspond to the initial conditions for the Hubble parameter and its first
derivative, we can solve the three following coefficients as
\ba
h_3 & = & -\frac{1}{36\Ma^2}\LF\mu + 6h_1\Ma^2(\la+4h_1\Ma^2) \RF\,, \label{h_3} \\
h_5 & = & \frac{1}{720\Ma^6}\LF \la\mu + 2h_1(3\la^2+11\mu)\Ma^2+132\la h_1^2\Ma^4 + 384h_1^3\Ma^6 \RF\,, \\
h_7 & = & -\frac{1}{90720 \Ma^8}\Big[ \mu (3 \la^2 + 26 \mu) + 6 h_1 \la (3 \la^2 + 73 \mu) \Ma^2 \nonumber \\
& + &  12 h_1^2 (141 \la^2 + 242 \mu) \Ma^4 + 17424\la h_1^3 \Ma^6 + 39168 h_1^4 \Ma^8\Big]\,. \label{h_7}
\ea
From the structure of the series it is apparent that each coefficient $h_{2n+1}$ is uniquely determined as by the previous ones. The highest
order coefficient appears always linearly, guaranteeing a real solution. Let us then consider the Friedmann equation (\ref{a_rho}) to determine
the density $\rho_0$. The first three nontrivial constraints are
\ba
0 & = & 3 \la \mu - 6 c_0 \mu^2/\Ma^2 - 4(3 \la^2 \rho_0/M_p^2 + \mu )\rho_0/M_p^2 - 36 h_1^2 (\la - 2 c_0 \mu/\Ma^2)\Ma^4\,, \label{a_fe1}  \\
%0 & = & \mu^2 - 6 c_0 \la \mu^2/\Ma^2 - 12 \la^3 \rho_0/M_p^2 + 4 \la \mu \rho_0/M_p^2 - 36 h_1^2 \la (\la - 2 c_0 \mu/\Ma^2)\Ma^4\,, \label{a_fe1} \\
0 & = & (3\la^2-\mu)2h_1\rho_0/M_p^2 + 9h_1\LF \la - 2c_0\mu/\Ma^2\RF\left[ h_1\la+6\LF h_1^2+h_3\RF\Ma^2\right]\Ma^2\,, \label{a_fe2} \\
0 & = & -\rho_0(2h_1^2+h_3)\LF3\la^2-\mu\RF/M_p^2 \nonumber \\ & + &  {9 (\la-2 c_0 \mu/\Ma^2) \Ma^2 \left[3 \left(10 h_3 h_1^2+10 h_5 h_1 +
h_3^2\right)
   \Ma^2+2h_1 h_3 \la\right]} \label{a_fe3} \,.
\ea
By construction, we expect all of these to be identically satisfied, once the solution to the ansatz equation is used in conjunction with one
of them to fix $\rho_0$. This is because the form of Friedmann equation was already dictated by the Bianchi identity given the ansatz we have.
Thus the redundant system (\ref{a_fe1}-\ref{a_fe3}) serves as a consistency check. The density we obtain is
\be
\rho_0 = \frac{3M_p^2 (\la - 2c_0\mu/\Ma^2) (\mu - 12 h_1^2 \Ma^4)}{12\la^2 - 4 \mu}\,.
\ee
The parameters of the model should be chosen such that $\rho>0$.

\subsection{Special case $\mu=3\la^2$}
\label{app2}

The hyperbolic cosine bounce $a(t) = a_0\cosh{\sqrt{\frac{\la}{2}}}t$ belongs to the class of models satisfying our ansatz, in particular when
$\mu=-6\la^2$. There are solutions of this
type and its generalizations admitted by the nonlocal field equations as we saw above. One special case which our treatment in (\ref{app1}) however does not cover is when $\mu=3\la^2$. Let us consider now this. Now the constraints (\ref{tr-cons1}-\ref{tr-cons3}) dictate we should have
\be
\frac{c_0\la}{\Ma^2}= \frac{1}{6}\,, \quad \La =-\frac{3}{4}\la M_p^2\,.
\ee
So the quadratic coefficient is fixed by $\la$, but now the propagator $\Ga(\la)$ is not determined by the trace equation and is left as free variable. We cannot now write the energy density as Eq.(\ref{a_rho}) but obtain instead
\be \label{a_ro2}
\frac{\rho}{M_p^2} = 3(1-\Ga(\la))\left[H^2+\frac{1}{\la}\LF 6H^2\dot{H}+2H\ddot{H}-\dot{H}^2\RF\right] - \frac{3}{4}\Ga(\la)\la\,.
\ee
The three first conditions from the ansatz equation (\ref{a_ans}) are equivalent to Eqs.(\ref{a_re1}-\ref{a_re3}). The first one can be now simplified to
\be
0 = \la^2 + 2 h_1 \la \Ma^2 + 4 (2 h_1^2 + 3 h_3) \Ma^4\,,
\ee
the rest remain the same. From the constant part of the Eq.(\ref{a_ro2}) we get the expression for radiation density as
\be
\rho_0=-\frac{3M_p^2(1-\Gamma(\la))(\la^2-4h_1^2\Ma^4)}{4\la}\,.
\ee
Again we can check that the constraints from the Friedmann equation expansion at higher order in $(\Ma t)$ are identically satisfied.

\subsection{Quadratic model}
\label{app4}

%{\it may be we can discuss this in the appendix}
To briefly check the consistency of the solution for the model (\ref{quad}) that includes only the near-trivial truncation of the theory,
we recall that in fourth order gravity the Friedmann equations can be written as
\ba
3f(\dot{H}+H^2)-\frac{1}{2}F & = & 3H\dot{f}-\rho\,, \label{q1} \\
f(\dot{H}+3H^2)-\frac{1}{2}F & = & \ddot{f}+2H\dot{f}+p\,, \label{q2}
\ea
where $f=dF/dR$. If we now plug in (\ref{quad}) and say $H(t)= h_1 \oa^2t + \dots$, at zeroth order the above equations yield $\rho=18M_p^2c_0 h_1^2\Ma^4/a^4$
and $p=6M_p^2c_0 h_1^2\Ma^4/a^4$, matching with our constraints. The expansion of $H$, which does not terminate, could be continued to higher orders. Thus solutions can be obtained,
but they are not cosmologically viable as shown in \ref{la<0}.

\subsection{Harmonic Ricci-spaces}

\label{app3}

Also the case of vanishing $\la$ is a special case apart from the previous ones considered in \ref{app1},\ref{app2}, since we have used the assumption $\la\neq 0$ there. Now the constraints (\ref{tr-cons1}-\ref{tr-cons3}) from the trace turn out to imply the following relations between the model parameters
\be
\frac{c_1}{\Ma^4}= \frac{1}{2\mu}\,, \quad \Lambda = -\frac{3\mu c_0}{2\Ma^2} - \frac{\mu^2c_2}{2\Ma^6}\,, \quad c_1=0 \,.
\ee
These are impossible to satisfy. So in our case we must have nonzero $\la$. However, a different set-up can be considered when the trace equation is $T = -R$ as usually and not identically satisfied.
The simplest example is $\mu=\la=0$. In these spacetimes the solutions of GR and the nonlocal model coincide exactly. Perturbations can of course behave completely differently, and therefore the stability of these spaces would be interesting to study. However, since these spaces are not bouncing we don't pursue this further here.

\end{document}